\documentclass{aastex62}

\received{}
\revised{}
\accepted{\today}
\submitjournal{ApJ}

\shorttitle{Mass of SNe II progenitors}
\shortauthors{Straniero et al.}

\begin{document}

\title{The initial mass-final luminosity relation of type II supernova
progenitors. Hints of new physics?}

\correspondingauthor{Oscar Straniero}
\email{oscar.straniero@inaf.it}

\author[0000-0002-5514-6125]{Oscar Straniero}
\affiliation{Istituto Nazionale di Astrofisica, Osservatorio d'Abruzzo,
Via Maggini s.n.c., 64100, Teramo, IT}
\affiliation{Istituto Nazionale di Fisica Nucleare, Laboratori Nazionali del Gran Sasso, IT}

\author{Inma Dominguez}
\affiliation{Departamento de Fisica Teorica y del Cosmos, Universidad de Granada,
18071 Granada, ES}

\author{Luciano Piersanti}
\affiliation{Istituto Nazionale di Astrofisica, Osservatorio d'Abruzzo,
Via Maggini s.n.c., 64100, Teramo, IT}

\author{Maurizio Giannotti}
\affiliation{Physical Sciences, Barry University, 11300 NE 2nd Ave., Miami Shores, FL 33161, USA}

\author{Alessandro Mirizzi}
\affiliation{Dipartemento Interateneo di Fisica ``Michelangelo Merlin'', Via Amendola 173, 70126 Bari, IT}
\affiliation{Istituto Nazionale di Fisica Nucleare, Sezione di Bari, IT}



\begin{abstract}
We revise the theoretical initial mass-final luminosity relation for progenitors of type IIP and IIL 
supernovae.
The effects of the major uncertainties, as those due to the treatment of convection, 
semiconvection, rotation, 
mass loss, nuclear reaction rates and neutrinos production rates are discussed in some details. 

The effects of mass transfer between components of close-binary systems are also considered.
By comparing the theoretical predictions to a sample of type II supernovae for which 
the initial mass of the progenitors and the pre-explosive luminosity are available, 
we conclude that stellar rotation may explain a few progenitors 
which appear brighter than 
expected in case of non-rotating models.  In the most extreme case, SN2012ec, an initial rotational velocity up to 300 km s$^{-1}$ is required. 
Alternatively, these objects could be mass-loosing components of close binaries.
However, most of the observed progenitors appear fainter than expected.
This occurrence seems to indicate that the Compton and pair neutrino energy-loss rates,  
as predicted by the standard electro-weak theory, are not efficient enough and that an additional 
negative contribution to the stellar energy balance is required.   
We show that axions coupled with parameters accessible to currently planned experiments, such as IAXO and, possibly, Baby-IAXO and ALPS II, may account for the missing contribution to the stellar energy-loss. 

\end{abstract}


\section{Introduction}
The progenitors of type II supernovae (SNe) are massive stars that retain part of their H-rich envelope 
up to the onset of the core collapse. Recent surveys devoted to the search of type II progenitors, 
found that the majority of the supernovae of this class observed in the local Universe, those of type IIP 
(plateau) and type IIL (linear),
 are produced by red supergiants (RSG) whose luminosity never exceeds $\log L/L_\odot=5.1$ \citep[see the recent review by][]{smartt2015}. 
According to current massive star models, such a pre-explosive luminosity corresponds to an initial mass\footnote{
With {\it initial mass} we intend the mass at the zero-age-main-sequence (ZAMS) or, equivalently, 
the mass at the beginning of the H-burning phase.}   
of about 16 - 18 M$_\odot$, with some differences from author to author, depending on the adopted treatment of 
convective mixing and/or the initial rotational velocity \citep{limongi2000,woosley2002,hirschi2004,ET2004,wh2007,georgy2013,farmer2016,limongi2018}.  
This occurrence is at odd with an early theoretical suggestion for which a type II supernova is expected to 
be the final fate of red supergiants with initial mass up to $\sim 30$ M$_\odot$ \citep[e.g.,][]{wh2007}. 
Several solutions of this puzzling problem have been proposed \citep{smartt2015}, among which 
observational biases or  higher mass-loss rate for the more massive and brighter progenitors, 
which eventually explode as SNe of other types. 
Alternatively, it is possible that for stars with initial mass $M> 18$ M$_\odot$  the core collapse is not 
followed by a SN explosion. In these {\it failed-supernova} scenario, 
the energy deposited by neutrinos 
may be not enough to sustain the forward shock, thus leading to the formation of a black hole.   
Recent parametric studies of core-collapse models have investigated the conditions for 
which a star can (or can not) 
explode \citep{oconnor2011,ugliano2012,horiuchi2014,pejcha2015,sukhbold2016,ertl2016,ebinger2019}.
They found that there is not a single mass below which all stars explode.
Rather, there are islands of
``explodability'', separated by regions of  progenitor masses corresponding to non-exploding models.
In general, it appears that 95\% of type II supernovae should have initial masses $9 < M/M_\odot < 22$. 
Successful explosions are obtained for the majority of the models within this mass range.
Type II supernovae with initial mass between 25 and 27 M$_\odot$ are also possible, while 
models in the range 22 and 25 M$_\odot$ and between 27 and 30 M$_\odot$ in most 
cases collapse into black holes, directly or by fall back.  
 Eventually,
stars whose initial mass is above 30 M$_\odot$ lose their H-rich envelope before the final collapse
and may give rise to type Ib or Ic supernovae\footnote{In case of mass loss driven by Roche-lobe overflow in a close binary system, the mass of a SNe Ib/Ic progenitor can be smaller}.  
In summary, the picture arising from these theoretical studies is much more challenging and rich 
than the simple
scenario derived from the few observations of RSG progenitors available so far. Nonetheless, 
the lack of RSG progenitors more massive than 18 M$_\odot$ still appears in 
conflict with the theoretical expectation. 

Two alternative methods to estimate the progenitor masses have been recently developed 
\citep{utrobin2009,spiro2014,jerkstrand2014,barbarino2015,valenti2016,morozova2018}. The first
 is based on the best 
fit of the observed SN light curve, which provides information about the mass of the 
exploding stars and of the
dense circumstellar material supposed to be the result of the progenitor stellar wind.  
According to \citet{morozova2018}, the progenitor masses estimated with this method
range between 10.9 and 22.9 M$_\odot$ (95\% C.L.) and are consistent with a Salpeter's mass distribution. 
The upper mass bound, in particular, is substantially larger than that obtained 
by means of the pre-explosive luminosities and in a better agreement with the theoretical expectations.
The second method makes use of late-time spectra (nebular phase) to measure the 
oxygen yield of the supernova. Oxygen is a product of the hydrostatic nucleosynthesis and depends 
on the progenitor mass. Also in this case, the resulting masses are generally larger than those 
estimated by means of the pre-explosive luminosity \citep[see Figure 7 in][and  section 4]{davies2018}.     

Motivated by this mass discrepancy,
in this work we have investigated possible revisions of the current scenario of massive star evolution 
that might imply a variation of the pre-explosive luminosity of red supergiant progenitors 
of type II SNe. 
In the next section we revise the standard scenario, discussing the uncertainties 
affecting massive star models with mass ranging between 11 and 30 M$_\odot$. 
Then, in section \ref{sec_axions} we discuss a possible \textit{new-physics} solution.
Specifically, we show that axions or axion-like particles (ALPs), whose couplings
with photons and electrons are compatible with current bounds and 
 accessible to the next generation of experiments, 
could be efficiently produced in the stellar cores of massive stars, thus 
leading to an increase of the energy-loss rate. 
A summary of our conclusions follows. 
Numerical expressions for the axion rates used in this work and some additional discussion about 
their relevance for different plasma conditions are illustrated in the appendix.

\section{Standard models}\label{sec_standard}
In this section we review the current massive star models and the physical processes which 
determine their final luminosity. We will make use of models computed by other groups and available 
in the extant literature, and of a new set of models
specially computed by means of the Full Network Stellar evolution code 
\citep[FuNS,][]{straniero2006,piersanti2013}. Let us first summarize the main features of this code.

\subsection{The FuNS code}
The solver we use to integrate the differential equations describing the 
stellar equilibrium structure and the chemical evolution is based on a classical Henyey method.
It was originally derived from the former FRANEC code \citep{chieffistraniero1989}. 
The last version of the FuNS allows us to choose different degrees of coupling between 
the equations describing the physical structure 
and those describing the chemical evolution, as due to nuclear burning and mixing. For the models discussed in this work, we have adopted 
a scheme in which  the stellar structure equations (i.e., hydrostatic equilibrium, mass continuity, energy conservation and energy transport) 
and the chemical evolution equations due to the nuclear burning are solved simultaneously. 
Then, the stellar zones which are thermally unstable are separately mixed. 
When dealing with the advanced stages of the evolution of a massive star, such a scheme is a good compromise between accuracy, velocity and stability of the numerical solutions. 
Our code also allows three different choices of the time dependent mixing scheme, namely: diffusive, advective or the non-local algorithm described in \citet{straniero2006}. For the purpose of this paper the three schemes are equivalent.
In order to calculate the temperature gradient and the average turbulent velocity (or diffusion coefficient) 
in the convective regions we use, as usual, the mixing length theory. In particular, we follow the scheme  
described in \citet{CG}. The mixing length parameter ($\alpha_{ML}$) has been calibrated 
by means of a Standard Solar Model \citep{piersanti2007}. Note that the value of $\alpha_{ML}$ depends 
on the adopted solar composition.
Since we adopt the \citet{lodders2009} composition, we found $\alpha_{ML}=1.9$.

In general, the convective boundaries are fixed according to the Ledoux criterion (see section \ref{sec_semico}). In addition, during the He-burning phase, the external border of the convective core is found by imposing the condition of marginal stability, i.e., $\nabla_{rad}$ strictly equal to  $\nabla_{ad}$ \citep[for more details see][]{straniero2003}. 
Except for the models described in section \ref{sec_overshoot}, no convective overshoot is applied.

The FuNS code also account for stellar rotation \citep{piersanti2013}. 
As usual, we assume shellular rotation, i.e., constant angular velocity and composition on the isobaric 
surfaces \citep[see,e.g.,][]{maeder_book}. 
As illustrated in section \ref{sec_rotation}, the most important instabilities induced by rotation, which are responsible for transport of angular momentum and mixing, are considered.
The potential effects of the magnetic field on rotation are ignored.

The FuNS has been optimized to handle large nuclear networks, as those required 
to follow the neutron-capture nucleosynthesis in AGB stars \citep[see][]{straniero2006}. 
However,
a minimal nuclear network that includes all the reactions providing  
the major contribution to the energy production is enough for the purpose of the present paper
and allows to reduce the computational time consumption. 
Hence, we have considered 33 isotopes, from $^1$H to $^{56}$Ni, coupled by 19 reactions 
for the H burning 
(those describing a full pp-chain and CNO-cycle) and an $\alpha$-chain of 51 reactions,
 plus the $^{12}$C$+^{12}$C, $^{12}$C$+^{16}$O and $^{16}$O$+^{16}$O, for the more advanced burnings. 
 The reaction rates are from the STARLIB database \citep{starlib}. 
 Most of the H and He burning reaction rates in this database are based on available experimental data. Rate estimates of reactions for which no experimental information exists, or extrapolation  to low temperatures at which no experimental rates exist, are obtained by means of  statistical  (Hauser–Feshbach)  models  of  nuclear reactions, computed using the code TALYS \citep{talys2008}.  
 Finally, electron screening corrections are computed according to \citet{dewitt1973,graboske1973,itoh1979}.
 Other details on the FuNS input physics are here provided:

\begin{itemize}
  \item Neutrino Energy Loss: 
     \begin{itemize}
        \item Plasma $-$ \citet{haft1994}
        \item Photo $-$  \citet{itoh1996a}  
        \item Pair  $-$  \citet{itoh1996a}  
        \item Bremsstrahlung $-$ \citet{dicus1976}
        \item Recombination $-$ \citet{BPS1967}
     \end{itemize} 
  \item Radiative Opacity:    
      \begin{itemize}
         \item \citet{AF1994}+\citet{OPAL}+\citet{LAOL}
      \end{itemize}       
  \item Electron Conductivity: 
      \begin{itemize}
        \item \citet{potekhin1999a,potekhin1999b}
      \end{itemize}
  \item Equation of State:
     \begin{itemize} 
          \item \citet{opaleos} + \citet{straniero1988} \citep[see also][]{prada2002}
     \end{itemize}
  \item Mass Loss:
      \begin{itemize} 
        \item \citet{dejager1990} 
      \end{itemize}
   \item External Boundary Conditions:
       \begin{itemize}
           \item  scaled solar $T(\tau)$ relation \citep{KS1966}
        \end{itemize}
\end{itemize}
 
As it is well known, the luminosity of a red supergiant and its He-core mass are closely related. 
Indeed, the mass of the core determines the physical conditions 
(T and $\rho$) of the H-burning shell, which is the source of energy that sustains the shining of these stars. 
After the He-burning phase, the central temperature rapidly increases becoming larger than $\sim 5\times 10^8$ K. 
At that temperatures, the production of neutrinos, as due to the Compton scattering and $e^+e^-$ annihilation,
 becomes very efficient. 
 As a result, the advanced stages of the evolution of a massive star 
 are controlled by the neutrino energy loss that largely 
 overcomes the photon energy loss. Such an occurrence causes a rapid drop of the evolutionary time scale, 
 which becomes much smaller than the H-burning time scale. Hence, 
 since the exhaustion of the central carbon and until the final collapse, the He-core mass and, in turn, the
 stellar luminosity stop to grow. 
In the rest of this section we will review the most important physical processes 
that determine the growth of the He-core mass up to this constant value 
and, hence, their impact on the pre-explosive luminosity.

The new models here presented have masses between 11 and 30 M$_\odot$. If not differently
specified, we have assumed solar composition, i.e., Z=0.014 and Y=0.27.
All the models, except those with $M=11$ M$_\odot$, have been evolved up to the Si burning, 
when the central temperature is $\sim4\times 10^9$ K. 
Instead the calculation of the $M=11$ M$_\odot$ models has been stopped just 
after the off-center Ne ignition.

\begin{figure}
    \centering
	\includegraphics[width=14cm,keepaspectratio]{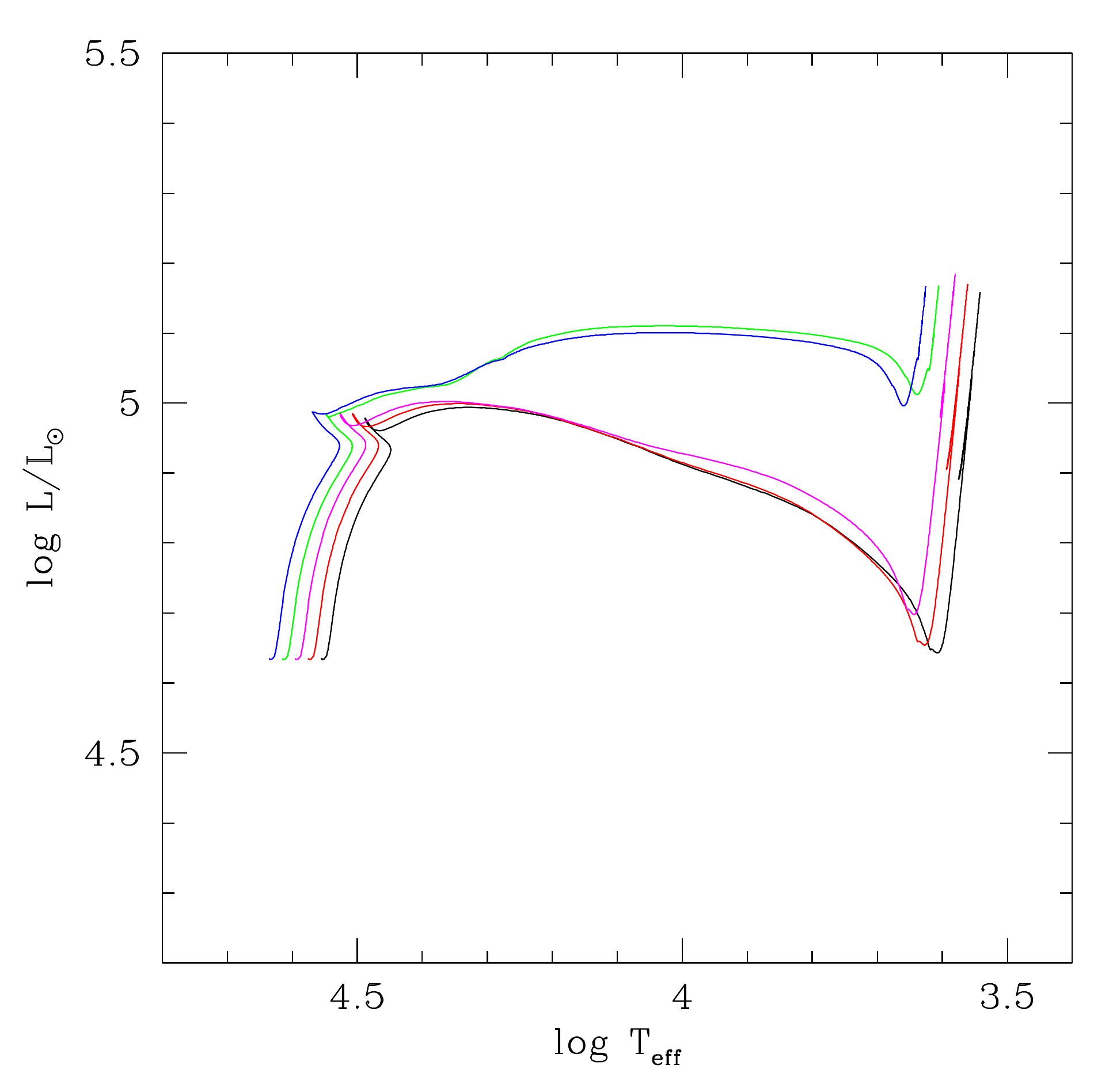}
    \caption{Evolutionary tracks of 20 M$_\odot$ models as obtained under
     different assumptions about the mixing efficiency in the semiconvective zones (see text for details):
      $\eta=10^{-3}$ (green), $\eta=10^{-4}$ (blue), $\eta=10^{-5}$ (magenta), $\eta=10^{-6}$ (red), $\eta=0$ (black). 
      Note that, in order to distinguish the different models,
      relative shifts of $\Delta \log T_{eff} = 0.02$ have been applied to the tracks with $\eta>0$.}. 
    \label{semico}
\end{figure}

\begin{figure}
	\includegraphics[width=\columnwidth]{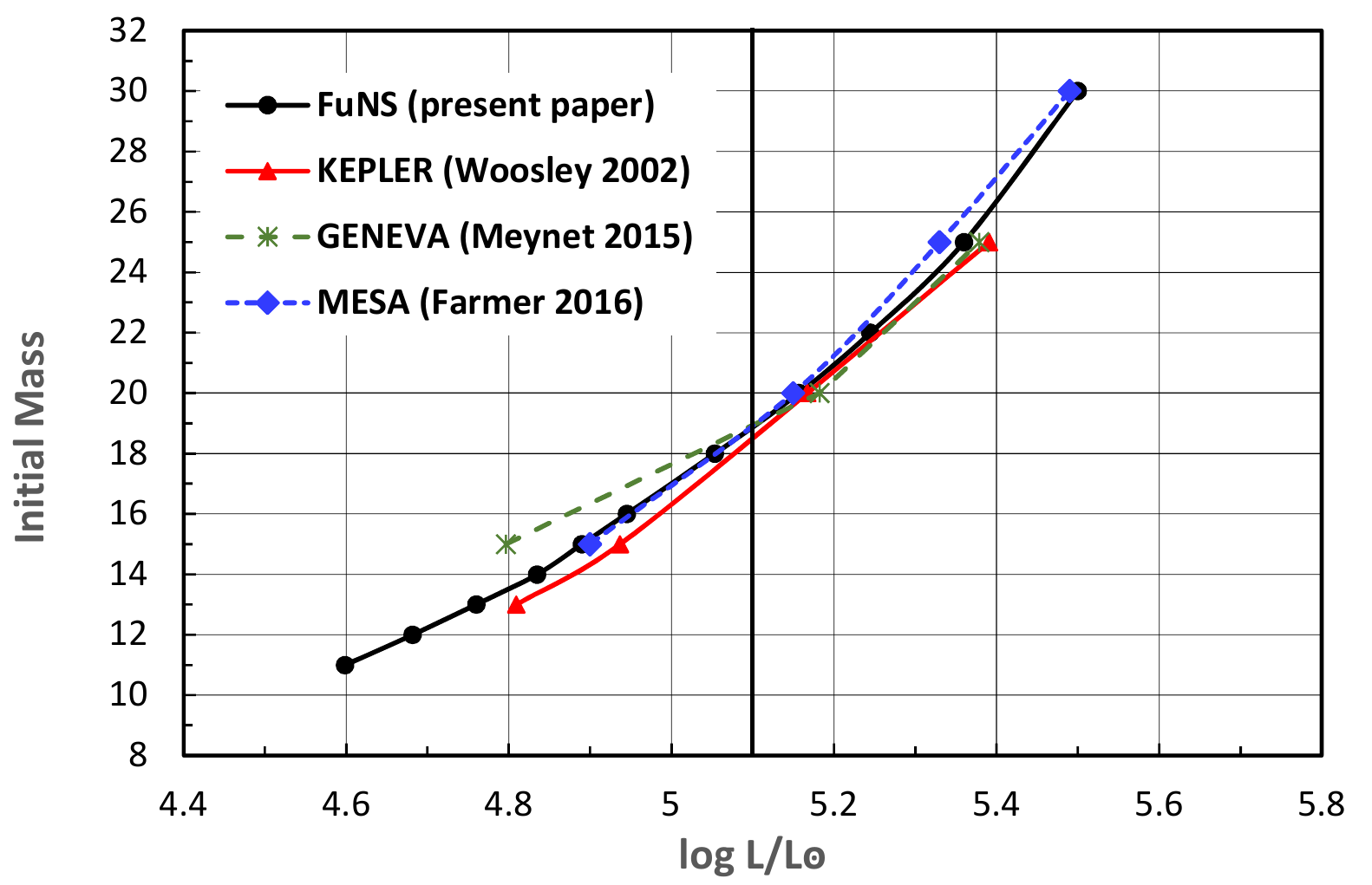}
    \caption{Initial mass-final luminosity relation for non-rotating stellar models. At the onset of the core collapse, all the models are red supergiants, except for the 25 M$_\odot$ GENEVA model.}
    \label{imflstd}
\end{figure}

\begin{table}
	\centering
	\caption{Initial Mass of type II SN progenitors with final luminosity
    $\log L/L_\odot=5.1$.}
	\label{tab1}
	\begin{tabular}{lcl} 
		\hline
		 Name & M/M$_\odot$ & Reference\\
		\hline
          \multicolumn{3}{c} {Standard non-rotating Models} \\
        \hline
		FuNS & 18.9 & present work\\
		KEPLER & 18.5 & \citet{woosley2002}\\
		\hline
        \multicolumn{3}{c}{with convective overshoot}\\
        \hline
        $\beta = 0.01$ & 17.0 & present work\\
        $\beta = 0.02$ & 13.8 & present work\\
        GENEVA & 18.9 & \citet{meynet2015}\\
        MESA  & 18.9 &  \citet{farmer2016}   \\
        ORFEO & 17.4 & \citet{limongi2018}\\
        STARS & 15.6 & \citet{ET2004}\\
       \hline
       \multicolumn{3}{c}{with different compositions} \\
       \hline
       Y$=$0.32 Z$=$0.014 & 17.6 & present work \\
       Y$=$0.27 Z$=$0.006 & 17.7 & present work \\
       \hline
       \multicolumn{3}{c}{with rotation} \\
       \hline
       $v_{ini}=200$ km s$^{-1}$ & 11.8 & present work \\
       $v_{ini}=150$ km s$^{-1}$ & 13.0\footnote{It is a blue supergiant} & \citet{limongi2018} \\
       $v_{ini}=0.4 v_{crit}$ & 17.1 & \citet{meynet2015}\\
       \hline
       \multicolumn{3}{c}{with axion energy loss} \\
       \hline
       $g_{10}=0.6$, $g_{13}=0$ & 19.8 & present work \\
       \hline
       $g_{10}=0.6$, $g_{13}=4$ & 20.9 & present work \\
       \hline
\end{tabular}

\end{table}

\subsection{Semiconvection}\label{sec_semico}
Among the principal uncertainties affecting H and He burning models of massive stars, those related to the treatment of turbulent mixing induced by convection still remain the most debated. In this context, a longstanding problem concerns the criterion adopted to establish if a not homogeneous zone is unstable against convection. In particular, in case of a negative molecular weight gradient, it may happens that a thermally  unstable  stratification  is  stabilized  against
adiabatic convection by a gradient in composition.
 This phenomenon, often called semiconvection, is particularly relevant for massive stars \citep[see][]{langer198}. 
 In this case, the widely adopted Schwarzschild criterion, which ignores the stabilizing effect of the molecular 
 weight gradient, appears inadequate to model the mixing and the consequent heat transport.
  On the other hand, the Ledoux criterion, which includes composition effects, hampers mixing in semiconvective 
  regions. 
 
Concerning massive stars, semiconvective zones firstly appear during the H-burning phase, 
when the convective core
 recedes, as a consequence of the conversion of H into He, and a region of negative $\mu$ gradient is left outside. 
If according to the Scharzschild criterion these layers are efficiently mixed, 
a prompt He ignition occurs after the exhaustion of the central hydrogen, when 
the star is still a compact blue supergiant. Then, during most of the He-burning phase the star remains hot 
and only towards
the end of this phase it eventually becomes a RSG. 
On the contrary, if according to the Ledoux criterion the mixing is inhibited in the semiconvective layers,
after the central-H exhaustion the stellar envelope expands  
and the star directly evolves 
into a cool and red supergiant. In this case the He ignition takes place on the RSG branch. 

A simple analysis of stability \citep[see][]{KW1990} demonstrates that even if the normal convective modes are stable, oscillatory modes can exists that may induce some mixing in the semiconvective layers. 
On the other hand, observations of massive stars in the Milky Way and in the Magellanic Clouds clearly 
favor a prompt evolution to the red supergiant phase \citep{stothers1992a,stothers1994}. 
Figure \ref{semico} shows some evolutionary tracks of the models we obtained by varying the damping factor 
($\eta$) of the semiconvective velocity ($v_{sc}$) with respect to the fully convective velocity ($v_{c}$). 
In practice, we have assumed that $v_{sc}=\eta v_{c}$ and varied the $\eta$ parameter from 0 to $10^{-3}$. 
Note that the case $\eta=0$ corresponds to the bare Ledoux criterion, while for $\eta=1$, the effect 
of the molecular weight gradient is fully suppressed.
Only models with $\eta \leq 10^{-5}$ become red supergiants before the He ignition. Note, however,
that  the variation of the final luminosity is rather small, namely $\Delta \log L < 0.03$.
In the rest of the paper, we will adopt the Ledoux criterion. 
The resulting initial mass-final luminosity relation is reported in Figure \ref{imflstd}. 
For comparisons, other relations, as  obtained from non-rotating models with moderate or no overshoot,  are also shown, namely KEPLER \citep{woosley2002}, GENEVA \citep{meynet2015} and MESA \citep{farmer2016}.
The small differences are likely due to the adopted treatments of overshoot, semiconvection and mass loss. 
The vertical line marks the observed upper limit for the luminosity of type II SN progenitors 
\citep{smartt2015}. The corresponding masses are reported in Table \ref{tab1}.

\subsection{Convective Overshoot}\label{sec_overshoot}
\begin{figure}
	\includegraphics[width=\columnwidth]{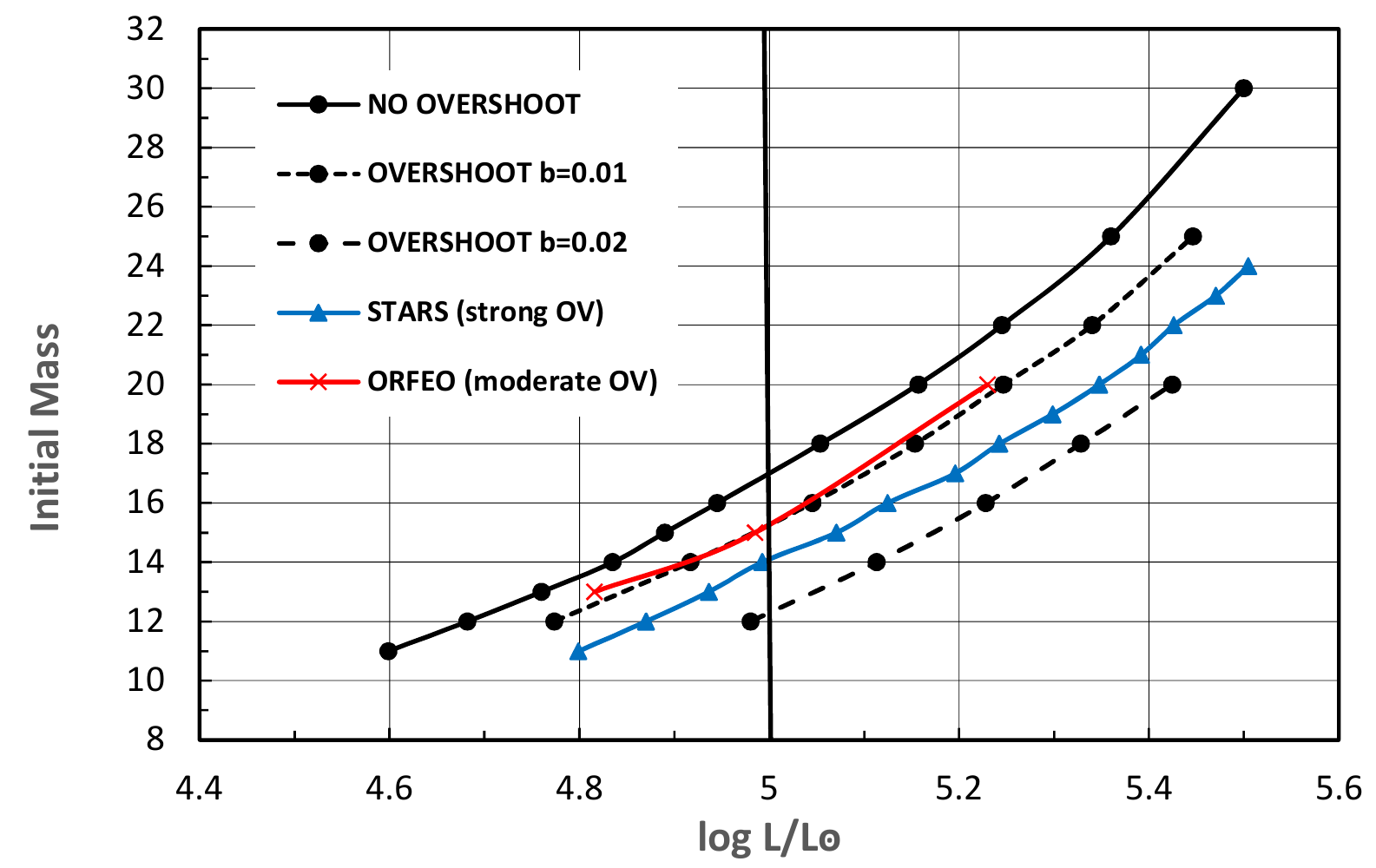}
    \caption{Initial mass-final luminosity relation from FuNS models with convective overshoot. For comparisons,
the FuNS standard relation (no overshoot) and those derived  from models of other authors are also shown 
(see text for references).}
    \label{overshoot}
\end{figure}

The so-called convective-core overshoot is another phenomenon that may affect the final He-core mass and, in turn, the final luminosity of a massive star. Indeed, owing to the inertial motion 
  of the convective cells, a transition zone, rather than a sharp discontinuity,  should exist 
  between stable and unstable regions. In principle, the extension of this zone and the mixing efficiency within it depend 
  on the convective velocity at the boundary layer and on the steepness of the entropy gradient. 
  Therefore,  a unique prescription to be applied to all the convective boundaries appears unrealistic. In most cases hints on the overshooting extension may be provided by the observations of the stellar properties affected by this phenomenon. Concerning massive stars, \citet{stothers1992b} \citep[but see also][]{MM1989} have shown that the convective-core overshoot cannot be too large, otherwise it would be in  conflict with observations. Nevertheless, in order to quantify the effect of this phenomenon on the initial mass-final luminosity relation, we have computed two additional  set of models with convective-core overshoot. In practice, according to current hydrodynamical simulations of stellar convection \citep[see, e.g.,][]{freytag1996}, we have extended the mixing outside the fully convective core of the H-burning models by imposing an exponential decline of the convective velocity, namely:
\begin{equation}
      v_{ov}=v_{cct} \exp \left ( -\frac{\delta r}{\beta H_p} \right )
 \end{equation}
\noindent
where $v_{cct}$ is the convective velocity at the top of the fully convective core\footnote{This is the marginal stability border of the convective core, 
where $\nabla_{rad}=\nabla_{ad}$}, $\delta r$ is the radial distance from the convective border and $H_p$ is the pressure scale height. $\beta$ is a free parameter.
The results are shown in Figure \ref{overshoot} for $\beta = 0.01$ and 0.02, short- and long-dashed lines, respectively. 
For a given stellar mass, the final He-core mass is larger in case of convective overshoot and, in turn, the pre-explosive models are brighter than those obtained 
by neglecting the overshoot mixing (solid line).  
For comparison, we have also reported the initial mass-final luminosity relations from two extant set of stellar models 
with overshoot, namely: ORFEO \citep{limongi2018} and STARS (\citet{ET2004}\footnote{As derived from Figure 4 in \citet{smartt2015}).}. 
Note that a moderate overshoot (of the order of $0.1 H_p$) has been adopted in the ORFEO models, while the extra-mixing zone is rather extended in STAR models ($\sim 0.5 H_p$).
The corresponding upper mass bounds for the observed progenitors of type II supernovae are also listed in Table \ref{tab1}.
In case of large overshoot, this maximum mass 
would be particularly small, thus increasing the tension with the masses estimated by means of light curve
best fitting or from the oxygen yields measured in the nebular phase. \citep{morozova2018, davies2018}. 
In addition, due to the higher luminosity, the mass-loss is larger in models with convective-core overshoot. As a result, also the maximum mass of collapsing red supergiants is smaller than that predicted by models without overshoot.

\subsection{Rotation}\label{sec_rotation}
Rotational velocity of the order of 100-200 km s$^{-1}$ are commonly observed in main sequence massive stars 
\citep{hunter2008}. 
As it is well known, 1D hydrostatic codes may easily account for shellular rotation \citep{endal1976}.
Two are the major consequences of rotation, namely: i) the lifting due to the centrifugal force, and ii) the meridional circulation   
that arises because of the deviations from thermal equilibrium occurring in rotating structures (Von Zeipel's paradox).   
The lifting effect of rotation implies more expanded and cooler stellar structures. 
In addition, the mixing induced by  meridional circulation modifies the  mean molecular weight and the opacity of the envelope.
As a result, larger convective core are expected in H and He burning stars with rotation. 
However, a quantitative estimation of these effects requires a reliable description of the angular momentum redistribution 
as due to both dynamical and secular instabilities. 
Moreover, angular momentum loss driven by stellar wind, which is particularly intense 
when the star is a red supergiant, must be considered.
  
\begin{figure}
    \centering
	\includegraphics[width=14cm,keepaspectratio]{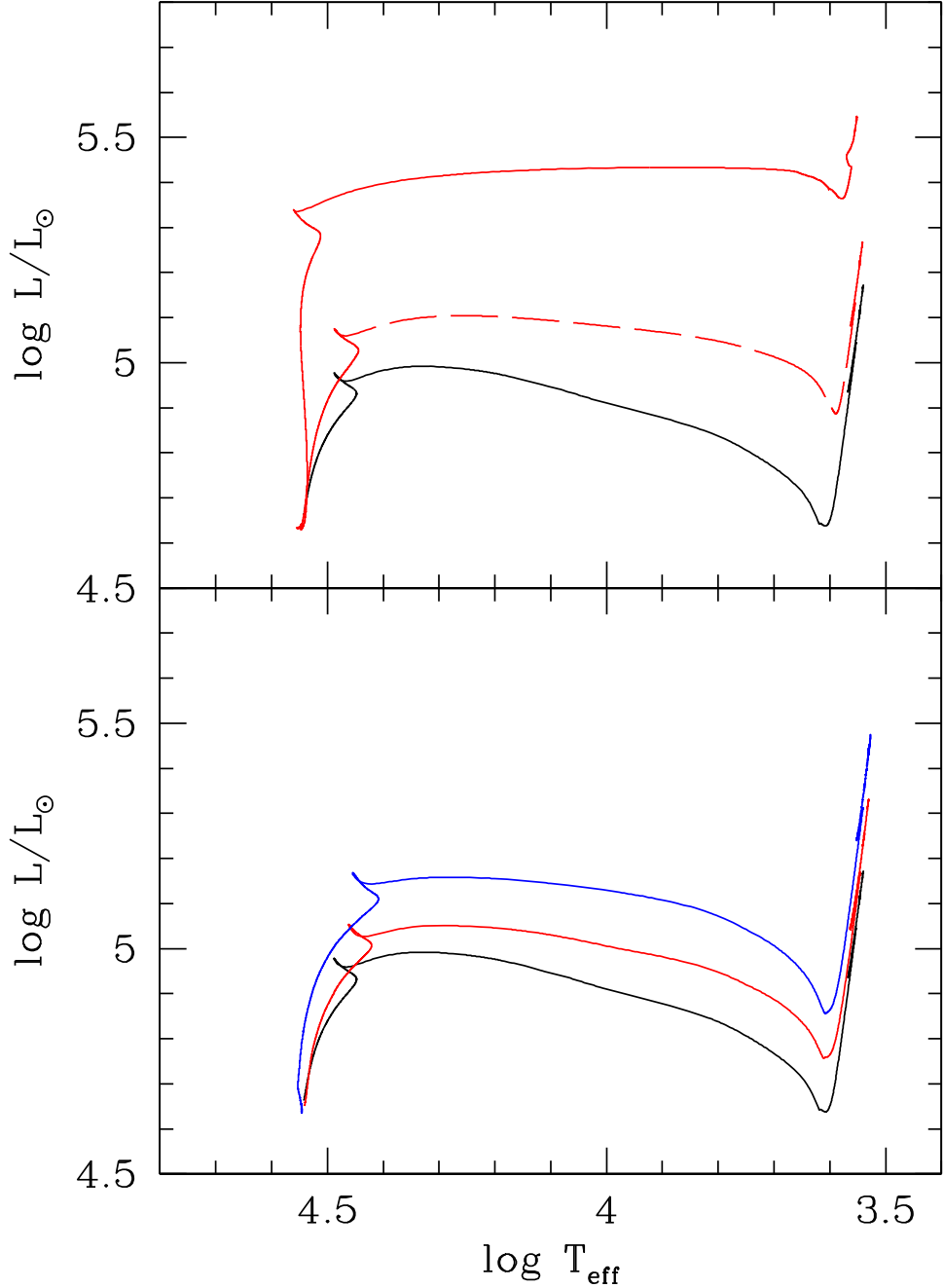}
    \caption{Evolutionary tracks of rotating models (M=20 M$_\odot$, no mass loss) for different sets of 
    $v_{ini}$, $f_c$, $f_\mu$. Upper panel: 100, 0.04, 0 (red-dashed line);
  200, 0.04, 0 (red-solid line). Lower panel: 200, 1, 0.01 (red-solid line); 200, 1, 0.005 (blue-solid line). 
  For comparisons, the non-rotating track is also reported (the black line in both panels).   
}
    \label{cmdrot}
\end{figure}

In the present version of the FuNS code the transport of angular momentum is treated as a diffusive process 
\citep[see details in][]{piersanti2013}. The most important instabilities induced by rotation 
are considered. 
The Eddington-Sweet (ES) and, to a less extend, 
the Goffrei-Shubert-Fricke (GSF) instabilities may produce significant effects 
on the evolution of massive stars \citep{heger2000}. 
According to \citet{KW1990} the ES circulation velocity is given by:
\begin{equation}
v_{ES}=\frac{\nabla_{ad}}{\delta(\nabla_{ad}-\nabla)} \frac{\omega^2 r^2 L}{(Gm)^2} \left ( \frac{2\epsilon r^2}{L} 
- \frac{2r^2}{m} - \frac{3}{4 \pi r^2 \rho} \right )
\end{equation}
where, $\omega$ is the angular velocity, $\epsilon$ is the rate of energy 
loss per unit mass, $\delta=-\frac{\partial \ln \rho}{\partial \ln T}$, $\nabla$ and $\nabla_{ad}$ are 
the temperature gradient and the adiabatic 
gradient, respectively. L, m, r, G and $\rho$  have the usual meaning.
As for convection, a negative molecular weight gradient reduces 
the efficiency of the meridional circulation. According to \citet{k74},
we model the effect of the $\mu$-gradient by
defining an equivalent $\mu$ current, which works against the ES circulation:
\begin{equation}
  v_\mu = f_\mu\frac{H_P}{\tau} \frac{\phi \nabla_\mu}{\nabla - \nabla_{ad}}
\end{equation}
where $H_P$ is the pressure scale height, $\phi= \frac{\partial \ln \rho}{\partial \ln \mu}$,   
$\nabla_\mu=\frac{\partial \ln \mu}{\partial \ln P}$
 is the $\mu$ gradient and $\tau$ is the thermal relaxation timescale (also called Kelvin-Helmholtz timescale).
$f_{\mu}$ is a (free) parameter used to tune the strength of the $\mu$ gradient barrier.
Hence, the effective circulation velocity is: $\left | v_{MC} \right |=\left | v_{ES} \right |-\left |v_\mu  \right |$. 
Such a velocity is used to calculate the angular momentum diffusion coefficient, while
a fraction of it, i.e., $v_{rot}=f_c v_{MC}$, is used to calculate the mixing induced by rotation. 
 In principle, both $f_\mu$ and $f_c$ should vary between 0 and 1.  These parameters could be calibrated
 by requiring that models reproduce some observables, such as the modification of the 
 surface composition due  to mixing induce by the meridional circulation 
 \citep[see, e.g.,][]{heger2000,limongi2018,maeder_book}.
Alternatively, asteroseismology may provide more stringent constraints to the efficiency of the angular momentum transport.
Indeed, it allows to measure core rotation rates of stars in different evolutionary phases \citep[see][ and references therein]{egg2019}.
The first results of the application of this method, mainly concerning low- and intermediate-mass stars, 
seem to indicate that the angular momentum redistribution is very efficient. 
In general, this occurrence should reduce the effects of rotation.  
Unfortunately, a clear and unique prescription valid for massive stars is still lacking. 
This problem represents a further uncertainty 
affecting massive star models.  
In Figure \ref{cmdrot} we report the evolutionary tracks of a 20 M$_\odot$ star as obtained under different assumption 
 about $f_c$, $f_\mu$ and the initial (uniform) rotational velocity ($v_{ini}$).
In general, the final He-core mass 
 is larger than that of non-rotating models. As a consequence, a rotating SN progenitor is
brighter than a non-rotating one. However, the extent of this 
effect depends on the assumed set of rotation parameters.  

\begin{figure}
	\includegraphics[width=\columnwidth]{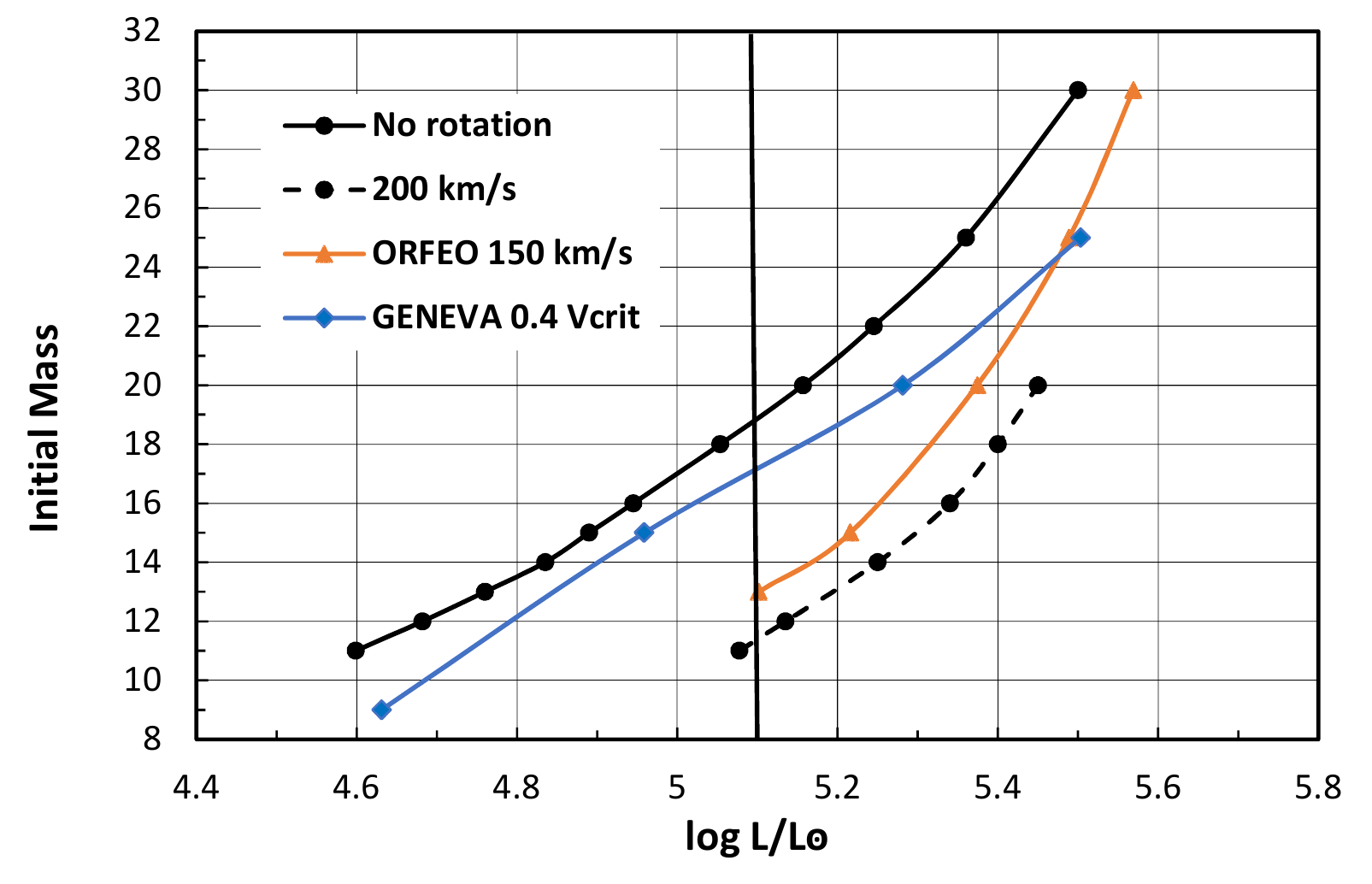}
    \caption{Initial mass-final luminosity relation for models with rotation (see text for references).
    Note that the FuNS rotating models with $M\ge 15$ M$_\odot$ leave the RSG branch before the occurrence of the core collapse. In contrast, none of the ORFEO models ends its evolution as red supergiant, while for the GENEVA rotating models the more massive pre-explosive RSG should have a mass between 15 and 20 M$_\odot$. 
    The vertical line at $\log L/L_\odot = 5.1$ marks the upper limit of the observed  pre-explosive luminosities \citep[according to][]{smartt2015}. }
    \label{rotation}
\end{figure}

In Figure \ref{rotation}, the variation of the initial mass-final luminosity relation for 
models with rotation is illustrated. 
The FuNS relation for $v_{ini}=200$ km s$^{-1}$, $f_c=0.04$ and $f_\mu=0$, is shown, together with
that from the ORFEO database \citep[][$v_{ini} = 150$ km s$^{-1}$, $f_c=1$ and $f_\mu=0.05$]{limongi2018}, and 
that from the GENEVA code \citep[][$v_{ini} = 0.4 v_{crit}$]{meynet2015}.
Note that a spread of $v_{ini}$ implies a spread of the final luminosity.
In other words, an observed pre-explosive luminosity do not correspond to a 
unique value of the initial mass. Instead a range of masse is possible, 
depending on the (unknown) initial rotational velocity. 

Another important consequence of rotation concerns the mass loss \citep{meynet2015}.
Indeed, owing to the higher luminosity, the mass-loss rate is larger in rotating models.
As a result, we find that rotating models ($v_{ini}=200$ km/s) whose initial mass is $M\ge 15$ M$_\odot$ leave the RSG branch before 
the final core collapse and, for this reason, they cannot be progenitors of type IIP/L SNe.

\subsection{Mass loss}\label{massloss}
The effect of the mass loss on the evolution of massive stars has been deeply investigated 
by several authors \citep[for a detailed analysis of the mass loss effects see][]{meynet2015}. 
The total amount of mass lost before the final collapse depends on the initial mass, the initial rotation velocity and the extension of the convective-core overshoot. 
For instance, basing on non-rotating FuNS models, 
a 13 M$_\odot$ star loses less than 2 M$_\odot$, while a 25 M$_\odot$ is expected to lose about 
half of its initial mass. In general, this theoretical prediction is in good agreement
with the observed mass ejecta of SN IIP progenitors \citep{morozova2018}.  
As mentioned in the previous section, the higher RSG luminosity developed in case of rotation 
implies a stronger mass-loss rate. For the same reason, also models with convective overshoot 
develop higher mass-loss rate. 
In general, the larger the mass-loss rate the lower the final luminosity. 
 We find, however, that the final luminosity does not depend on the mass lost during the RSG phase, but mostly on the amount of mass lost during the main sequence (see also section \ref{binary}). Therefore, since the mass-loss rate is much lower in MS than in the RSG branch, the sensitivity of the initial mass-final luminosity on the adopted mass-loss rate is rather small. For instance, we find that in a non-rotating model of 18 M$_\odot$ with mass loss, 
the final luminosity is only the 5\% lower than that of a model without mass loss. 

On the other hand, if the RSG mass loss is strong enough to erode the H-rich envelope, the model leaves the RSG branch prior to the final core collapse
and the resulting supernova cannot be of type IIP/IIL. This occurrence implies a maximum RSG mass-loss rate for a progenitor of these SN types.

In this context, recent evidences of late time enhanced mass loss in progenitors of some bright SNe IIP/IIL \citep{moriya2011,morozova2017,yaron2017} are not expected to modify the initial mass-final luminosity relation for these stars. However, to explain pre-explosive outbursts commonly observed in SNe IIn, \citet{fuller2017}  have recently suggested that gravity waves generated by  vigorous  convection  during  late-stage nuclear  burning could transport energy upward, causing a heating of the H-rich envelope and, in turn, an increase of the final luminosity. Although firm conclusions about the occurrence and, eventually, the strength of such a heating process in normal SN IIP progenitors have not  been established yet, we may exclude that it could  solve the discrepancy between the masses estimated from the SN light curve and those derived from the pre-explosive luminosity. On the contrary, this phenomenon would increase the tension. Indeed, the masses from light curves are usually larger than those estimated from the  initial mass-final luminosity relation (see the discussion in section \ref{sec:discussion}).

\subsection{Neutrinos}\label{neutrinos}
The rate of neutrino energy loss determines the time at which the evolutionary time scale becomes
 so short that the He-core mass stops to grow and a luminosity freeze-out occurs. 
In particular, a rate larger than that commonly estimated would imply fainter  SN II progenitors. 
The current rates are  computed on the base of the Weinberg-Salam electro-weak theory 
\citep[see, e.g.,][]{raffelt_book}. 
In the last 30 years, precision experiments have tested this quantum field theory at the level of 
one  percent  or  better. 
Note, however, that the neutrino rates depend on the Weinberg angle, whose value is not predicted by the electro-weak theory. In the present calculations we have used the value suggested by \citet{itoh1996a}, namely:  $\sin^2 \theta _W=0.2319$, which is very similar to that reported in the latest compilation of the Particle Data Group\footnote{http://pdg.lbl.gov/2019/reviews/rpp2018-rev-phys-constants.pdf} (0.23155). We have also verified that negligible variations of the final luminosity are obtained when the CODATA 2014 value \citep[0.2223,][]{codata} is adopted.
As a whole, the accuracy of the Compton and Pair neutrinos 
energy-loss rates, which are the dominant neutrinos processes in the present stellar model calculations, 
is better than 5\% \citep{itoh1996a,itoh1996b}. Such an uncertainty has little effects on the pre-explosive luminosity. 

On the other hand, some deviations from the standard electro-weak theory are not completely excluded. In particular, the existence of a 
non-zero neutrino magnetic moment
 would enhance the stellar energy-loss.
Experimental and astrophysical constraints provide upper bounds for this quantity. 
The more stringent experimental constraint is  
$< 2.9\time 10^{-11}$ $\mu_B$. It has been obtained by means of reactor 
neutrinos \citep[90\% C.L., ][]{GEMMA}. 
A more stringent astrophysical constraint, as based on the luminosity of the RGB tip 
of Galactic Globular Clusters, 
gives $< 2.6\time 10^{-12}$ $\mu_B$ \citep[68\% C.L., ][]{viaux2013_AA}.
The influence of a non-zero neutrino magnetic moment on the evolution of massive stars has been 
investigated by \citet{heger_giannotti}. They found 
that models of initial mass above 15 M$_\odot$ are practically insensitive to a neutrino magnetic 
moment lower than $10^{-11}$ $\mu_B$. However, they do not consider the
sensitivity of the final luminosity on the enhanced neutrino rate. Therefore, we have computed a model of 20 M$_\odot$ by assuming
$\mu_{\nu}=5\times 10^{-11}$ $\mu_B$ and we find that the resulting final luminosity is just a 2\% lower than that
of the corresponding $\mu_{\nu}=0$ model.     

\subsection{Nuclear reaction rates}\label{nuclear}
Similarly to neutrinos, also a variation of the C-burning temperature could anticipate or delay the time of the luminosity freeze-out. Indeed, owing to the strong sensitivity of the neutrino production rates on the temperature, a high 
C-burning temperature would anticipate the freeze-out, while  the opposite occurs in case of a low C-burning temperature. 
This temperature depends on the rate of the $^{12}$C$+^{12}$C reaction and on the 
amount of carbon left in the core after the He burning.

A large uncertainty affects the $^{12}$C$+^{12}$C reaction rate for $T\le1$ GK \citep[for a recent reanalysis,
 see][and references therein]{zickefose2018}. 
At present, direct measurements of the cross section of this reaction are only available for 
energies $>2.1$ MeV.  
Owing to a {\it molecular-like structure} of the $^{24}$Mg compound nucleus, 
unknown resonances are possible at lower energy. 
The existence of these molecular states  
would substantially enhance the low-energy cross section, thus reducing
the C-burning ignition temperature.  
In contrast, hindrance model \citep{jiang2007a} predicts a steep drop of the low-energy 
$S(E)$ factor\footnote{The $S(E)$ factor represents the pure nuclear contribution to the 
fusion cross section: $S(E)=\frac{\sigma(E)}{E \exp (2\pi\eta)}$, where $E$ is the energy, 
$\sigma(E)$ the total cross section and $\eta$ the Sommerfeld parameter.} and, hence,
would imply a larger C-burning ignition temperature. In this context, some hints may
be obtained with indirect measurements.
\citet{tumino} have recently presented new results  
based the so-called Trojan Horse Method (THM). 
The THM is an indirect method, which allows to measure the $S(E)$ factors 
of charge particle reactions down 
to astrophysical relevant energies, where data from direct methods are not available  
because of the strong Coulomb barrier and the consequent very small cross sections. 
In the range of energy of more interest to the C-burning stellar rate, Tumino et al. found  
a complex resonant pattern, but no evidence of hindrance.
However, the interpretation of these indirect measurements requires 
reliable theoretical models and accurate calibrations.
On this base, \citet{mukha2018} has argued that the THM overestimates the actual $S(E)$ factor of the $^{12}$C$+^{12}$C reaction. Therefore, more 
accurate experiments and theoretical 
models are needed to fix this important open issue of massive star evolution.
Meanwhile, we have checked the potential impact of the THM result by  
computing a 20 M$_\odot$ model with the Tumino et al. rate. We 
find that the final luminosity is only a 2\% higher than that obtained with 
the widely adopted \citet{cf88} rate. An illustration of the impact of the THM rate
on stellar models with lower mass, those that ignite C and Ne in degenerate conditions, 
can be found in \citet{nic2018}.

The C-burning rate also scales with the square of the carbon abundance. 
This carbon has been previously produced during the He-burning phase.
The reactions involved in the C production are: the
triple-$\alpha$, which represents the production channel, and the    
$^{12}$C$(\alpha,\gamma)^{16}$O, which represents the destruction channel.
While the rate of the triple-$\alpha$ reaction is known within a 10\% at the temperature of the He-burning in massive stars \citep{fymbo2005}, 
the  $^{12}$C$(\alpha,\gamma)^{16}$O reaction rate is still rather uncertain.
 The rate in the STARLIB compilation is based on the \citet{kunz2002} study. Recently, a complete reanalysis has been reported by \citet{deboer2017}. Basing on an R-matrix fit of all the available experimental data, they find a reaction rate slightly smaller than that of Kunz et al.. As a consequence a larger amount of C is expected in the center of a star at the end of the He burning phase. In Table \ref{tab-12CA}, we compare the results we obtain for a set of 20 M$_\odot$ models computed under different assumptions for the  $^{12}$C$(\alpha,\gamma)^{16}$O reaction rate. Within the uncertainty quoted by \citet{deboer2017}, we find a 22 \% variation of the central carbon mass fraction at the end of the He-burning phase.     
Such a variation of the C abundance produces sizable effects on the C burning \citep{imbriani2001,WW1993}. Indeed, a lower C abundance implies higher temperatures, smaller convective cores and, in turn, shorter C-burning lifetimes. It affects the compactness of the pre-explosive stellar structure as well as the hydrostatic and explosive nucleosynthesis. In principle, these occurrences could also affect the final luminosity.  However, as noted by \citet{lisbona}, 
a larger $^{12}$C$(\alpha,\gamma)^{16}$O rate also implies a larger He-burning lifetime. It occurs because the  $^{12}$C$(\alpha,\gamma)^{16}$O reaction consumes 1/3 of He fuel compared to the triple$-\alpha$, but release a similar amount of nuclear energy. As a consequence, the larger the  $^{12}$C$(\alpha,\gamma)^{16}$O reaction rate the slower the He consumption.
In practice, this variation of the He-burning lifetime compensates the effect of the variation of the central C abundance, so that a small variation of the final He-core mass is found. As reported in the third column of Table \ref{tab-12CA}, when the $^{12}$C$(\alpha,\gamma)^{16}$O is varied within the uncertainty quoted by deBoer et al., such an occurrence implies negligible variations ($<0.2$ \%) of the  final luminosity. 

\begin{table}
    \centering
    \caption{Central C mass fraction at the end of the He burning ($X_c$) and pre-explosive luminosity for 20 M$_\odot$ models computed with different  $^{12}$C$(\alpha,\gamma)^{16}$O reaction rates. Low and high refer to the lower and upper bounds quoted by  \citet{deboer2017}.}
    \begin{tabular}{lcc}
    \hline
       rate     & $X_c$   &   $\log L/L_\odot$\\
    \hline
        \citet{kunz2002}   &  0.285  & 5.1739 \\
        \citet{deboer2017}   &  0.309  & 5.1739 \\
         high   &  0.275  & 5.1740 \\
         low   &  0.342  & 5.1746 \\
     \hline
    \end{tabular}
    \label{tab-12CA}
\end{table}

\begin{figure}
	\includegraphics[width=\columnwidth]{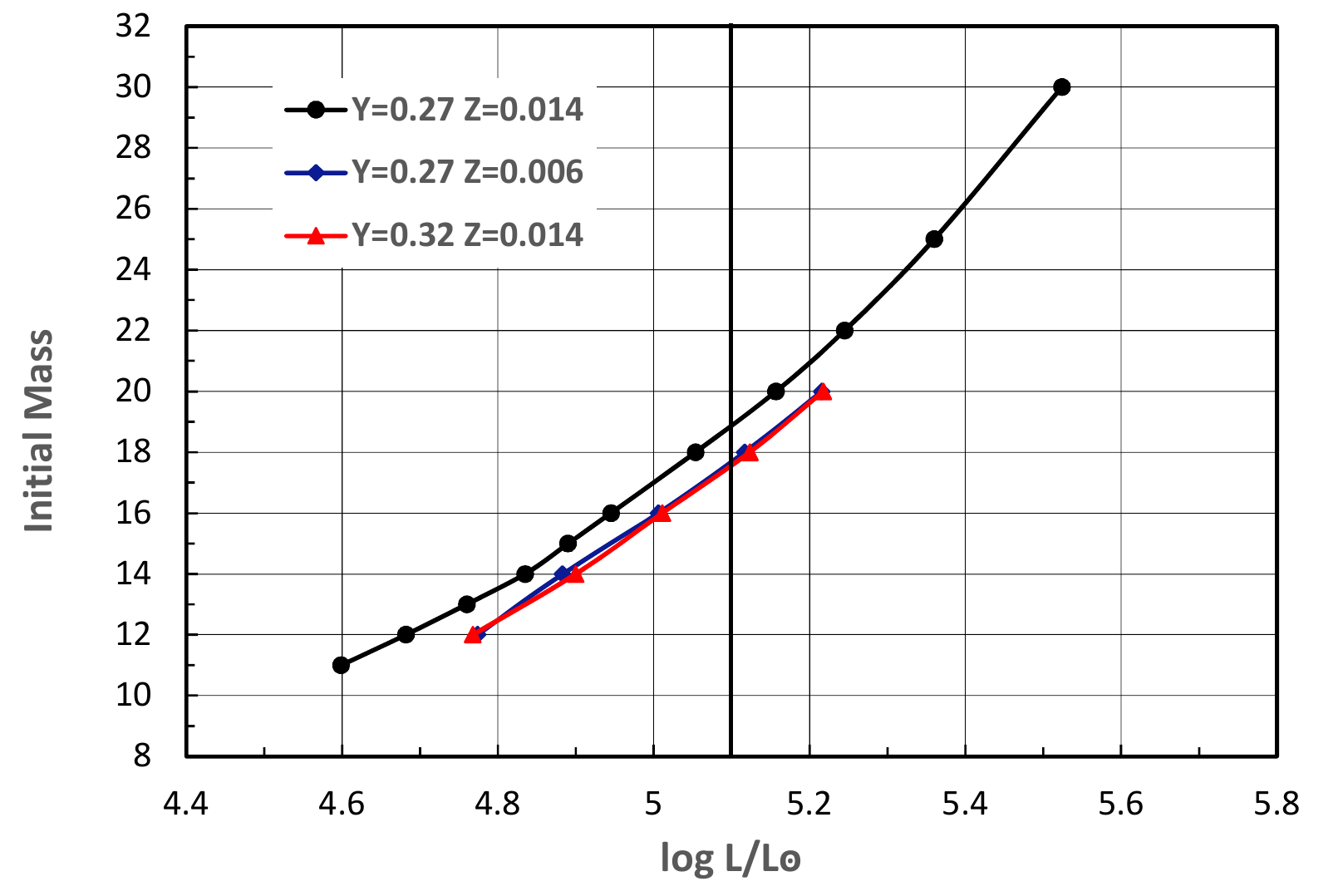}
    \caption{Initial mass-final luminosity relations for varied chemical compositions 
    are compared to the standard (solar composition). 
      See text for references.}
    \label{composition}
\end{figure}

\subsection{Chemical composition}\label{chemical}
Type II supernovae are found preferentially on the disks of spiral galaxies. 
Moreover, the host galaxies where the progenitors can be resolved belong to the local Universe. 
For these reasons, the parent stellar populations should 
be rather similar. 
Nevertheless a certain spread of the initial composition cannot be excluded.
Usually a solar composition is adopted. An increase of the initial He mass fraction and/or
a reduction of the initial metallicity both imply smaller envelope opacity and less
efficient H-burning. The resulting stellar structures are more compact and brighter. 
In Figure \ref{composition} we compare the initial mass-final luminosity relations 
obtained by increasing the 
initial He mass fraction, from Y=0.27 to Y=0.32, 
and by decreasing the initial metallicity, from Z=0.014 down to Z=0.006, to the standard one. 
The masses corresponding to a pre-explosive luminosity $\log L/L_{\odot}=5.1$ are reported in Table \ref{tab1}.

Summarizing, variations of Y and Z have opposite effects on the final luminosity.
Notice that a positive correlation between the He-mass fraction and the metallicity is a natural
consequence of the chemical evolution of galaxies \citep{pagel1998}. For this reason a spread of 
the initial composition should not produce a substantial modification of our findings.

\begin{figure}
	\includegraphics[width=\columnwidth]{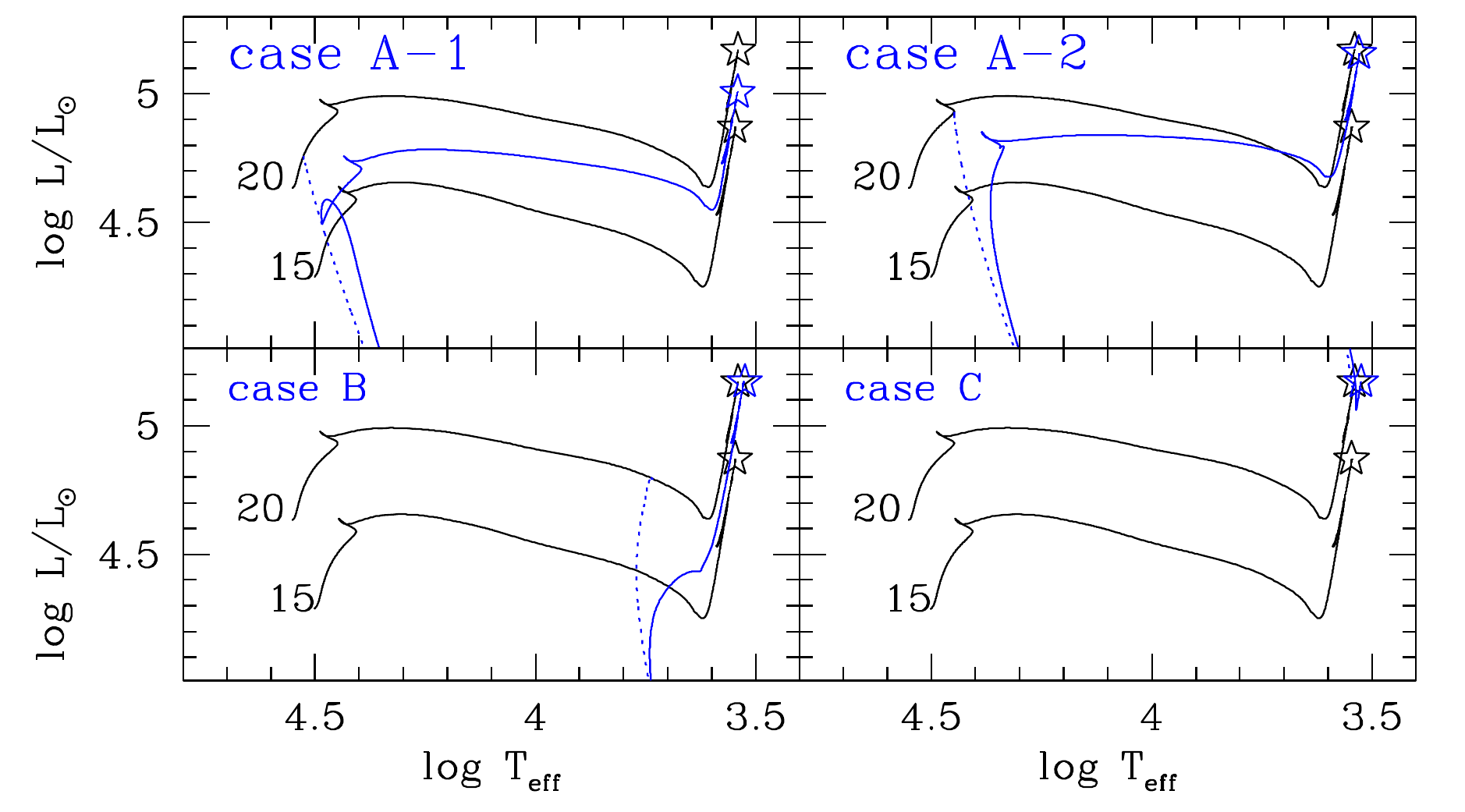}
    \caption{Effects of fast mass-loss episodes caused by Roche-lobe overflows along the evolutionary tracks of massive stars. The solid-blue lines in each panel show the evolution after a fast mass-loss episode of a star with initial mass of 20 M$_\odot$ that loses 5 M$_\odot$. The dotted-blue path corresponds to the mass-transfer phase. For comparison, we also report the evolutionary tracks of 15 and 20 M$_\odot$ single star models (solid-black lines). The asterisks indicate the final models.  The four cases, A-1, A-2, B and C, correspond to different choice of the epoch of the Roche-lobe overflow episode (see text for more details).}
    \label{abc}
\end{figure}

\subsection{Binaries}\label{binary}
More than 50\% of massive stars are in binary (or multiple) stellar systems and most  of these systems are close enough that one or more mass transfer episodes through Roche-lobe overflow may occur during the pre-supernova evolution \citep{sana2012}. In this case, not only mass, but also angular momentum is exchanged between the components of the binary. Tidal as well as dissipative forces may also induce a synchronization  of the stellar rotation rate and the orbital motion.  These occurrences are expected to modify the standard paradigm of massive stars, as based on models of single star evolution. Close-binary evolution may indeed produce a variety of different evolutionary channels \citep{paczynski1967, podsiadlowski1992, nomoto1995, vanbeveren1998, eldridge2008, yoon2010}. For this reason, the binarity is often invoked to explain peculiar objects or, more in general, to interpret the diversity of the core-collapse SN inventory. Famous examples are i) the anomalous SN1993J, whose progenitor was supposed to be a K-type supergiant hosted in a binary system with an  hotter companion, and ii) the SN1987A, whose progenitor  was clearly identified as a blue supergiant, possibly the secondary component of  a binary with initial mass ratio close to 1  \citep[for a review see][]{smartt2009}. 

 In general, three cases of mass transfer are usually distinguished, depending on the evolutionary status of the Roche lobe filling star. 
In case A, a star fills the Roche lobe during the main sequence, while in the B and C cases the mass exchange occurs when the donor is becoming or already is a red supergiant. Then, the mass lost by the donor flows through the internal Lagrangian point (L1) and eventually feeds an accretion disk around the smaller companion. 
It should be noted that main sequence massive stars are rather compact objects with radii of just a few R$_\odot$, so that the separation between the two components must be particularly short  for a case A Roche-lobe overflow. Such kind of binaries are probably a small fraction of all the binary systems harboring massive stars.  More frequent should be the other two cases, in which the Roche lobe filling star is a red or yellow supergiant. 

Among the various close binary evolutionary channels, here we are interested in those that at the end of the mutual interaction leave stars with a H-rich envelope whose mass is larger than about 2-3 M$_\odot$, so that they could be still considered as possible progenitors of type IIP SNe\footnote{A lower H-mass could be enough for the rare type IIL progenitors}.
Let us discuss the mass donors first. In order to investigate the consequences of an enhanced mass loss due to a binary interaction on the pre-explosive luminosity,  we have calculated a small set of 20 M$_\odot$ models, in which 5 M$_\odot$ of the H-rich envelope are artificially removed at a quite fast rate.  The fast mass-loss episode is switched on at different epochs. Then, when the total mass of 15 M$_\odot$ is attained, the mass loss is switched off and  the evolution has been continued up to the Si burning. No further mass transfer episode are considered during the remaining part of the evolution. In addition we ignore the possibility of a dynamical mass transfer with the consequent common envelope episode, a phenomenon that may possibly occur in case B and C of Roche-lobe overflow, due to the presence of an extended convective envelope in the donor star.  Nonetheless this simple experiment can provide a reasonable estimation of the effects of the binary interaction on the final luminosity. Four cases are here illustrated, namely: case A-1 and A-2, in which the fast mass loss takes place during the main sequence, when the central H-mass fractions are X=0.49 and X=0.04, respectively,  and the corresponding radii are 7 and 12 R$_\odot$; case B, in which the mass loss phase occurs when the star, after the main sequence, moves toward the RSG and  the radius attains 281 R$_\odot$; case C, in which the fast mass loss starts after the He-burning phase, when the radius attains $10^3$ R$_\odot$. The resulting evolutionary tracks are shown in Figure \ref{abc}. Note that in case A-2, B, and C the final luminosity practically coincides with that of a single star model with 20 M$_\odot$ (the initial mass value), while in case A-1, it is intermediate between that of the 15 and 20 M$_\odot$ single star models. In other words, the initial mass-final luminosity relation for the donor stars is insensitive to the mass transfer, provided that it occurs close to the end or after the main sequence. In all cases, if most of the mass lost by the donor is subtracted by the accretor (conservative case), the initial mass obtained by means of the initial mass-final luminosity relation for single stars  will be systematically larger than that estimated by means of the light curve fitting method. Among the 8 SNe progenitors whose mass has been derived with both methods, only one, i.e., SN2012ec, shows a similar behaviour, while the opposite is found for the others (see the discussion in section \ref{sec:discussion}).   

Concerning the accretors, also in this case the consequences of the mass transfer  depend on the evolutionary stage at the beginning of the accretion phase. According to \citet{podsiadlowski1992}, if the accretion occurs when the star is on the main sequence, the post-accretion evolution will be very similar to that of a single star whose mass is equal to the total mass, i.e., initial mass plus the accreted mass. On the other hand, when the accretion phase starts after the exhaustion of the central H, the pre-explosive structure will be a blue or yellow supergiant (see Figure 10 in \citet{podsiadlowski1992}). Likely, this is a consequence of the decrease of the core mass-total mass ratio ($\xi_c=\frac{M_c}{M} $) caused by the mass accretion process. Summarizing, only main sequence accretors may produce RSG progenitors and, possibly, generate a type IIP supernova. In this case the mass estimated by means of the initial mass-final luminosity relation for single stars progenitor should coincide with that obtained with the light curve fitting method. 

Finally, the possible gain of angular momentum, as due to the spin-orbit coupling, would have the same consequences illustrated in section \ref{sec_rotation}, namely, larger core masses, brighter progenitors and, in turn,  lower masses estimated by means of the initial mass-final luminosity relation.

\subsection{Numerical issues}\label{numerical}
The Henyey method employed in he majority of the stellar evolution codes is a first order implicit method and the accuracy of the solutions depends on the assumed mass and time resolutions. In the FuNS code, we adopt an adaptive algorithm to select the grid of mass shells. This algorithm controls the variations between adjacent mesh points of luminosity, pressure, temperature, mass, and radius. For instance, if A is one of $L, P, T, M$ or $R$ , it should be:
\begin{equation}
    0.8\times \delta < \frac{\left | A(N+1)-A(N-1) \right |}{A(N))}< \delta,  
\end{equation}
for each shell N.
In the present work we have used $\delta=0.05$ everywhere, except for the shells located around some critical points, such as the boundaries of the convective zones, for which we use $\delta=0.005$. This choice implies about 1000 mesh points for  main sequence models and up to 5000 for the most  advanced models. Recently, \citet{farmer2016} argue that a minimum mass resolution of 0.01 M$_\odot$ is necessary to achieve  convergence in the He-core mass within 5\%. However, as noted by \citet{sukhbold2018}, the effects of an improved resolution on the observable quantities, such as luminosity or effective temperature, are generally small. In particular, they found a weak trend to produce slightly larger He-core masses and, in turn, brighter pre-supernova models, when the zoning is finer.  
To check this occurrence, we have calculated a few additional evolutionary tracks with the further constraint that the difference in mass between two adjacent mesh points cannot exceeds 0.01 M$_\odot$. The resulting pre-explosive luminosity are less than the 1\% larger than that obtained under our standard choice for the grid resolution.   

\section{Beyond the  Standard Model: Axions}\label{sec_axions}
In this section we show that the emission of axions or axion like particles (ALPs) from the core of massive stars would alleviate the observed problems with the initial mass-final luminosity relation by shifting up 
the initial mass corresponding to the final luminosity $\log L/L_\odot = 5.1$.

The axion~\citep{weinberg1978,wilczek1978} is a light pseudoscalar particle predicted by the most widely accepted solution of the strong CP problem~\citep{PQ1977}
and a prominent dark matter candidate~\citep{Abbott:1982af,Dine:1982ah,Preskill:1982cy}.
Its interactions with photons and fermions ($f$) are described by the Lagrangian terms
\begin{eqnarray}
L_{\rm int}=- \frac{1}{4}g_{a\gamma} \, a F_{\mu \nu} \tilde F^{\mu \nu} - \sum_{f}g_{af} \, a \overline \psi_i \gamma_5 \psi_i \,,
\label{eq:Lint}
\end{eqnarray}
where $ F $ and $ \tilde{F} $ are the electromagnetic field and its dual while the $ \psi_i $ are the fermion fields.
In what follows we are interested in the coupling constants with photons, $ g_{a\gamma}= C_\gamma \alpha/2\pi f_a$, and with electrons $ g_{ae}= C_e m_e/f_a$, with $ C_\gamma$ and $ C_e $ model dependent parameters and $ f_a $ a phenomenological scale known as the Peccei-Quinn (PQ) symmetry breaking scale.
The PQ scale is not fixed by the theory and is, instead, constrained by experiments and through phenomenological considerations. 

\begin{figure}
	\includegraphics[width=\columnwidth]{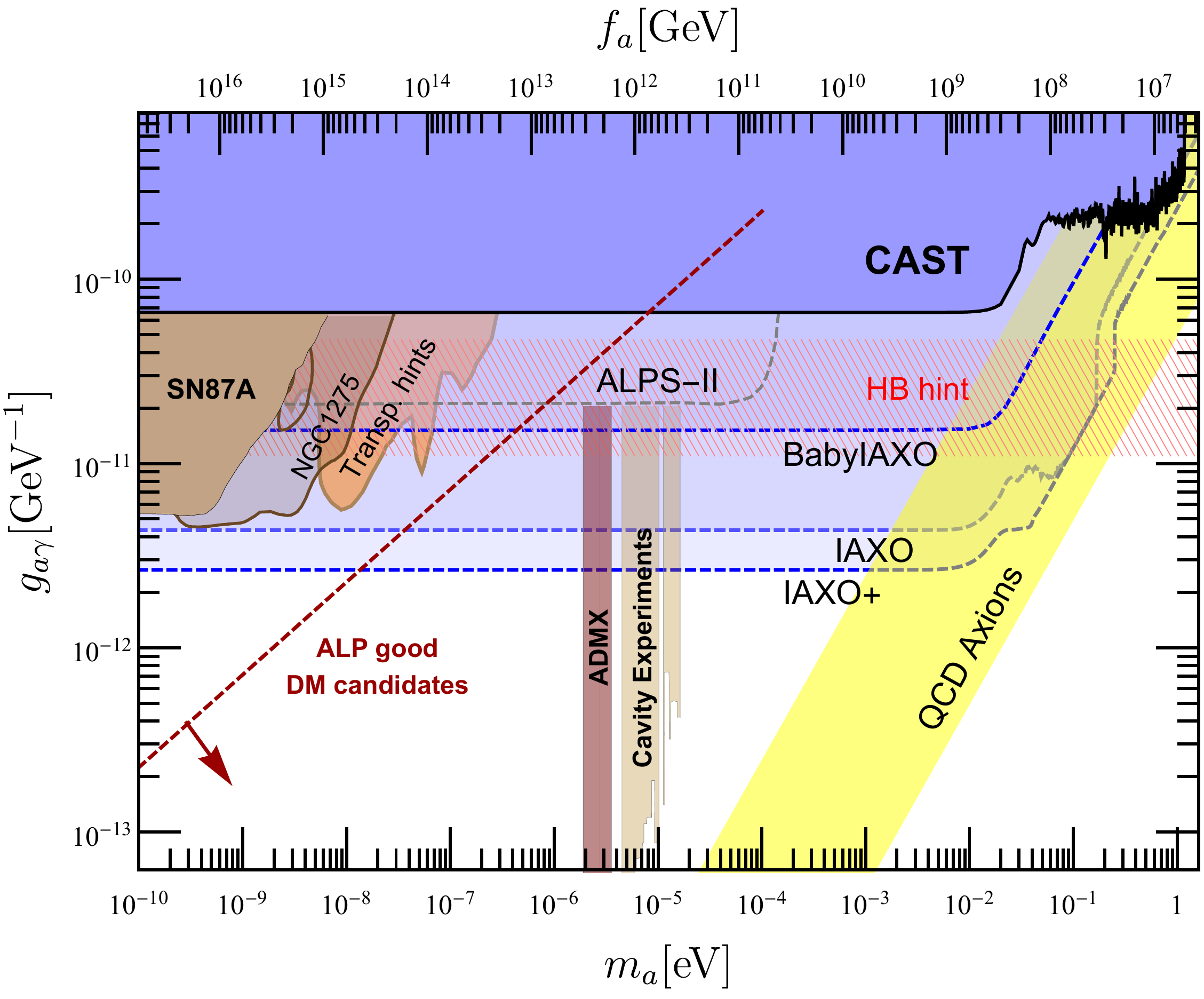}
	\caption{Overview of the axion and ALP  $ g_{a\gamma} $ vs. $ m_a $ parameter space, with axion hints and experiments~\citep{DiVecchia:2019ejf}. 
		The region below the dashed red line indicates the parameters that permit ALPs to be the totality of the cold dark matter in the universe~\citep{Arias:2012az}.
		Notice that the region hinted by HB stars is calculated for ALPs interacting only with photons~\citep{ayala2014,Straniero:2015nvc}. 
		See main text for more details.
		}
	\label{fig:gag_plot}
\end{figure}

Two minimal axion models are often considered as benchmarks in theoretical and  experimental studies: the Kim-Shifman-Vainshtein-Zakharov (KSVZ) model \citep{KSVZ1,KSVZ2} and the Dine-Fischler-Srednicki-Zhitnitsky (DFSZ) model \citep{DFSZ1,DFSZ2}. 
The most significant difference between the two models is in the couplings with electrons. 
The KSVZ, an example of hadronic axion, couples only to hadrons and photons at tree level.
Its coupling to electrons are therefore strongly suppressed ($ C_e\sim 10^{-3} $). 
DFSZ axions, on the other hand, couple also with electrons ($ C_e\sim 1 $). 
These QCD axion models share, however, the proportionality between the coupling constants, which measure the strength of their interactions, and the axion mass ($m_a$).
Hence, light axions are also weakly interactive. 
This is reflected in Figure \ref{fig:gag_plot} where the diagonal yellow band, representing the region where QCD axion models are expected to be found~\citep{DiLuzio:2016sbl}, shows the proportionality relation between the axion mass and its coupling to photons.

More general axion-like particles (ALPs), which do not satisfy specific relations between couplings and mass, 
are a common prediction of string theory and other theories of physics beyond the standard model~\citep{Witten:1984dg,Conlon:2006tq,Svrcek:2006yi,Arvanitaki:2009fg} though, in general, 
their existence is not related to the strong CP problem~\citep{Ringwald:2012hr}.
The proliferation of ALPs in extensions of the SM of particle interactions has motivated the experimental exploration of the ALP parameter space beyond the QCD axion band.

The recent years have seen an impressive experimental effort in the search of axions and ALPs~\citep[for a detailed review of the axion experimental landscape see][]{irastorza2018}. 
Among those, helioscopes, such as the CERN Axion Solar Telescope (CAST), which search for solar axions, are the best equipped to span large ALP mass regions.
The latest results form CAST~\citep{Anastassopoulos:2017ftl}, shown in blue in Figure \ref{fig:gag_plot}, have probed the axion-photon coupling down to $ g_{a\gamma} =0.66\times 10^{-10}$ 
GeV$ ^{-1} $, reaching the bound from globular cluster stars~\citep{ayala2014,Straniero:2015nvc}, 
in the mass region $ m_a \leq 0.02 $ eV. 
The next generation of axion helioscopes, BabyIAXO and IAXO~\citep{Armengaud:2014gea,Giannotti:2016drd,Armengaud:2019uso} are expected to push the limit on this coupling by a factor of, respectively,  $ \sim$3 and $ \sim $20, and to further probe the QCD axion band (see Figure \ref{fig:gag_plot}).
The IAXO+ sensitivity, also shown in the Figure, represents a possible improvement of the IAXO potential with upgraded components~\citep{Armengaud:2019uso}.
Axion haloscopes, such as the Axion Dark Matter eXperiment (ADMX), can reach considerable better sensitivity, though in very narrow mass bands. 
ADMX is already probing the QCD axion region~\citep{Du:2018uak} and the next generation of axion haloscopes will likely explore a large portion of the QCD axion band in the mass range between $ \sim 1\mu $eV and $ \sim 1 $ meV~\citep{Brubaker:2017rna,Brun:2019lyf,Alesini:2017ifp}.
At the same time, full laboratory experiments, such as ALPS II~\citep{Bahre:2013ywa}, allow for ALP searches in part of the unexplored ALP parameter space without relying on assumptions concerning the axion source.

The axion coupling to electrons is more difficult to probe experimentally.
The most stringent upper bounds have been obtained by the XENON100 collaboration \citep{xenon}, $g_{ae} < 7.7\times 10^{-12}$ (90 \% CL), LUX~\citep{Akerib:2017uem}, $g_{ae} < 3.5\times 10^{-12}$, and 
PandaX-II~\citep{Fu:2017lfc}, $ g_{ae}<4\times 10^{-12} $.
These bounds are, however, not yet competitive with the stellar bounds on this coupling. 

Besides dedicated terrestrial experiments, stellar observations offer a unique - and often very powerful - way to look at axions and other weakly interacting particles~\citep[see, e.g.,][]{raffelt_book,Raffelt:2006cw}.
Axions with masses below a few 10 keV can be efficiently produced in stellar hot interiors through 
thermal processes similar to those allowing the production of neutrinos. 
Once produced, these weakly interactive particles can freely escape providing a net sink of energy.
Considerations about stellar evolution have provided very strong bounds on the axion couplings to photons, electrons, and nucleons, often exceeding the results achieved in laboratory experiments. 
Currently, the strongest bounds on both the axion-photon and axion-electron coupling were derived by stellar evolution considerations. 
 The most stringent upper bound on the axion-photon coupling, 
$g_{a\gamma} < 0.65\times 10^{-10}$ GeV$^{-1}$ (95\% CL), was obtained by \citet{ayala2014}. 
It is based on the measurement of the so-called $R$ parameter, essentially the ratio of the
time scales of Horizontal Branch (HB) and Red Giant Branch (RGB) stars, in 32 galactic Globular Clusters. 
As previously recalled, this result 
was recently confirmed, for ALP masses below 20 meV, by the CAST collaboration. 
Analogously, the most stringent experimental upper bounds on the axion-electron coupling  were derived from astrophysical considerations. 
\citet{MB2014} \citep[see also][]{isern2018} report $g_{ae}<2.8\times 10^{-13}$, as obtained from their analysis of the  white dwarfs (WDs) luminosity functions, \citet{corsico2016} find $g_{ae}<7\times 10^{-13}$, from the period drift of pulsating WDs, and \citet{viaux2013} report $g_{ae}<4.3\times 10^{-13}$, from the luminosity of the RGB tip of the Globular Cluster M5. 
These constraints can be improved using multi-band photometry of multiple globular clusters~\citep[see, e.g.,][]{Straniero:2018fbv}.

More intriguingly, a series of astrophysical observations have also shown an excessive energy loss in many stellar systems, which could be accounted for by additional light, weakly interactive particles~\citep[for a recent review see][]{Giannotti:2016drd}. 
The \textit{new-physics} interpretation of these anomalies, from observations of WDs, RGB, and HB stars, has resulted in the selection of axions and ALPs, among the various light, weakly interacting particles, as the only candidates that can explain all the excesses~\citep{Giannotti:2015dwa,Giannotti:2015kwo}.
Given the very different stellar systems in which excessive energy losses have been observed, it is quite remarkable that one single candidate can explain all the anomalies.
The ALP parameter range invoked to explain the anomalous observations (at 2$ \sigma $) is shown in the purple hashed region in Figure \ref{fig:sensitivity}.
The analysis indicates a preference for ALPs coupled to both electrons and photons, though a vanishing photon coupling would still be compatible with the observations within 1$ \sigma $.
The electron coupling, on the other hand, is predicted to be finite with a $ \sim 3\sigma $ statistical significance.
In any case, the hints point to a well defined area in the axion parameter space which is in part accessible to the next generation of axion probes, including BabyIAXO, IAXO and ALPS II. 

\subsection{Axion impact on massive stars}
As discussed above, the enormous majority of results on the impact of axions on stellar evolution involves low mass stars, that is stars of mass similar or smaller than the mass of the sun. 
Part of the reason is that low-mass stars are the most abundant constituents of the galactic stellar populations, allowing a quite larger statistics. 
However, the energy loss rate via axions is very sensitive to the temperature (see the appendix) 
and this favors the hotter environment of the core of more massive stars.  
In spite of that, only a few attempts have been made to study the axion effects on 
the evolution of stars a few times heavier than the sun. 
\citet{Dominguez:1999gg} studied the impact of axions on the Asymptotic Giant Branch (AGB)
 phase of low and intermediate mass stars (up to 9$ M_{\odot} $), showing that 
they produce relevant changes into the evolution and, in particular, 
cause a deficit of massive C-O white dwarfs.
Later, \citet{Friedland:2012hj} considered the impact of axions on the He-burning phase
 of solar metallicity stars with mass in the 8-12$ M_{\odot} $ range, 
focusing, in particular, on the axion effects on the blue-loops.
In the present work, we are considering the impact of the axion emission 
on the evolution of massive stars and show how axions may 
reduce the pre-explosive luminosity of SN type II progenitors and,
in turn, increase the estimated mass.

The axion luminosity is given by:
\begin{equation}
L_a=\int_{0}^{R}4\pi r^2\rho \epsilon_a dr
\end{equation}
where R is the stellar radius, $\rho(r)$ and $T(r)$ the density and the temperature internal profiles, and $\epsilon_a (\rho,T)$ is
the rate of energy drained by axions per unit mass (see appendix). 
In the hot plasma of a star, $ \epsilon_a $ takes contribution mostly from the following processes:
\begin{enumerate}
\item Primakoff, i.e., photon-axion conversion in the electrostatic field of electrons and ions;
\item Compton, i.e., scattering of photons on electrons or ions;  
\item electron-positron annihilation $e^+e^-\to \gamma + a$, also known as pair production;
\item bremsstrahlung $ e +Ze\to  e + Ze +a $ on electrons and nucleons.
\end{enumerate}
The numerical recipes we have used to estimate these rates are summarized in the appendix, so the results can be reproduced and tested in future analyses. 

\begin{figure}
	\includegraphics[width=\columnwidth]{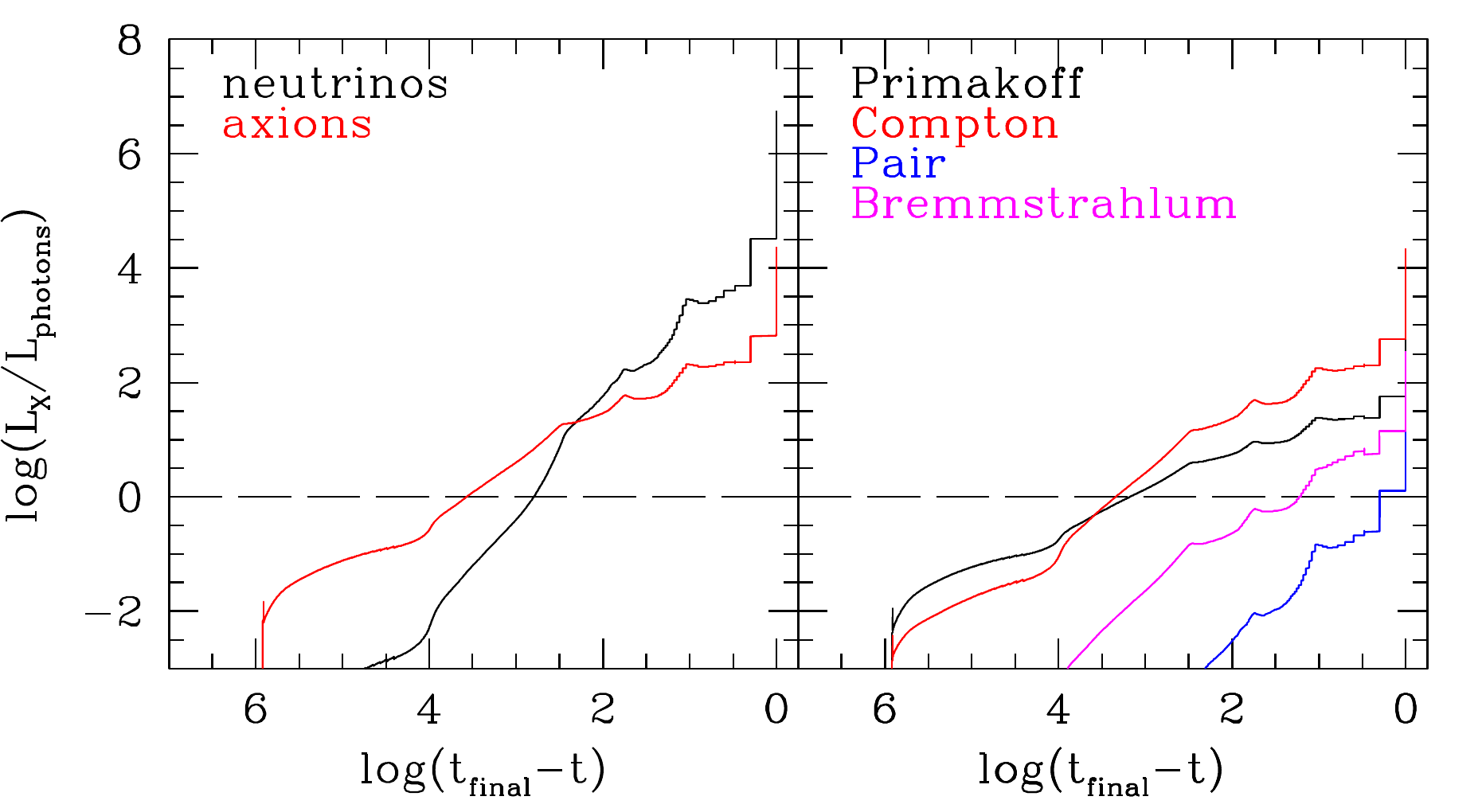}
    \caption{Left: the axion luminosity is compared to the neutrino luminosity along the evolution of a 20 M$_\odot$ model; Right: for the same model, the Primakoff, pair, bremsstrahlung and Compton contributions to the total axion luminosity are shown. The adopted values of the the axion-photon and axion-electron couplings are $g_{ae}=4 \times 10^{-13}$ and $g_{a\gamma} =0.6 \times 10^{-10}$ GeV$^{-1}$, roughly corresponding to the available experimental and astrophysical upper bounds. The evolutionary time is in yr.}
    \label{aloss}
\end{figure}

\begin{figure}
   \centering
	\includegraphics[width=15cm,keepaspectratio]{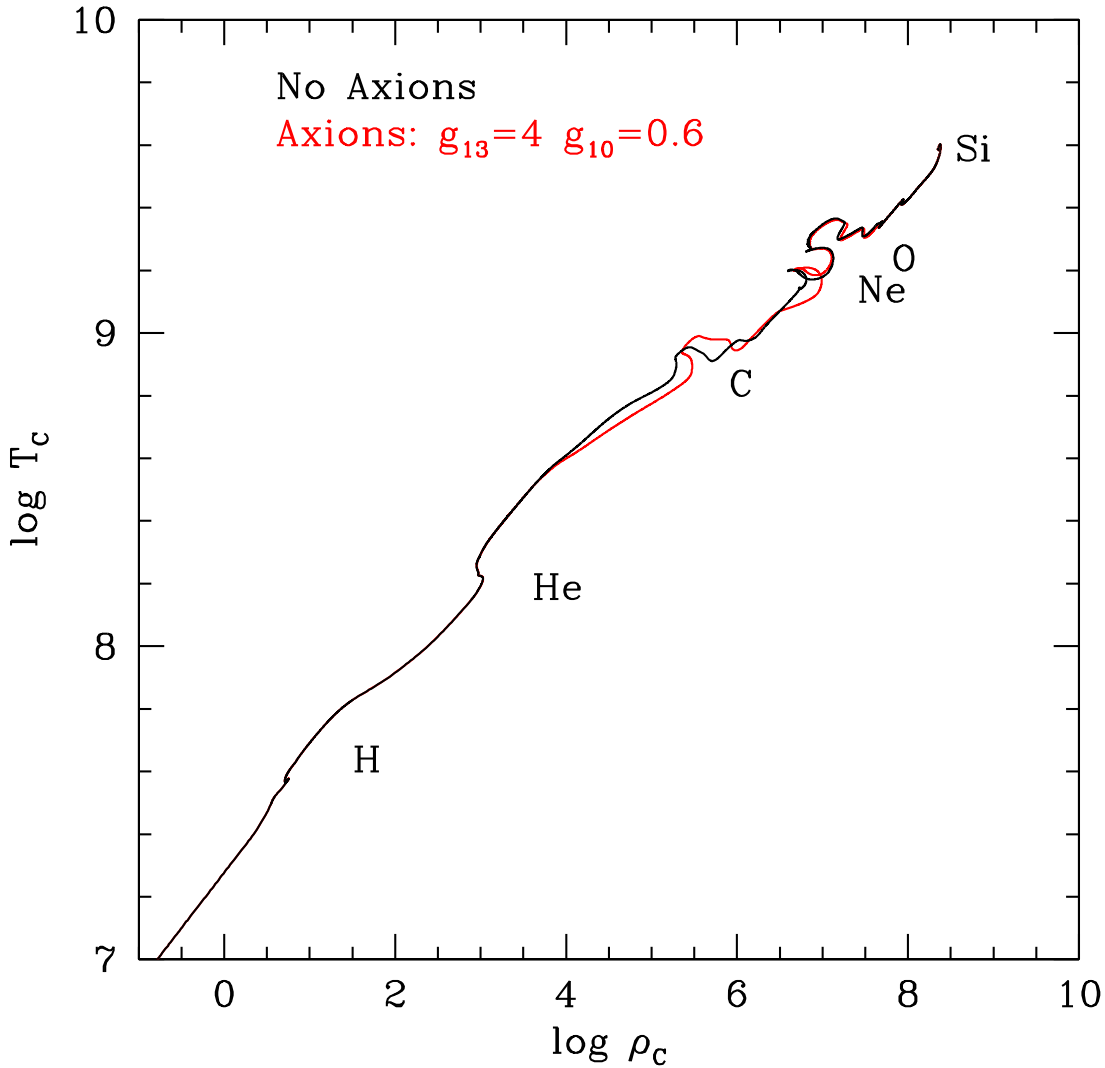}
    \caption{Central temperature versus central density for two 20 M$_\odot$ models: standard (no axions), black line, and non standard ($g_{10}=g_{a\gamma}/10^{-10}$ GeV$^{-1}=0.6$ and $g_{13}=g_{ae}/10^{-13}=4$), red line. }
    \label{tcroc}
\end{figure}

\begin{figure}
    \centering
	\includegraphics[width=14cm,keepaspectratio]{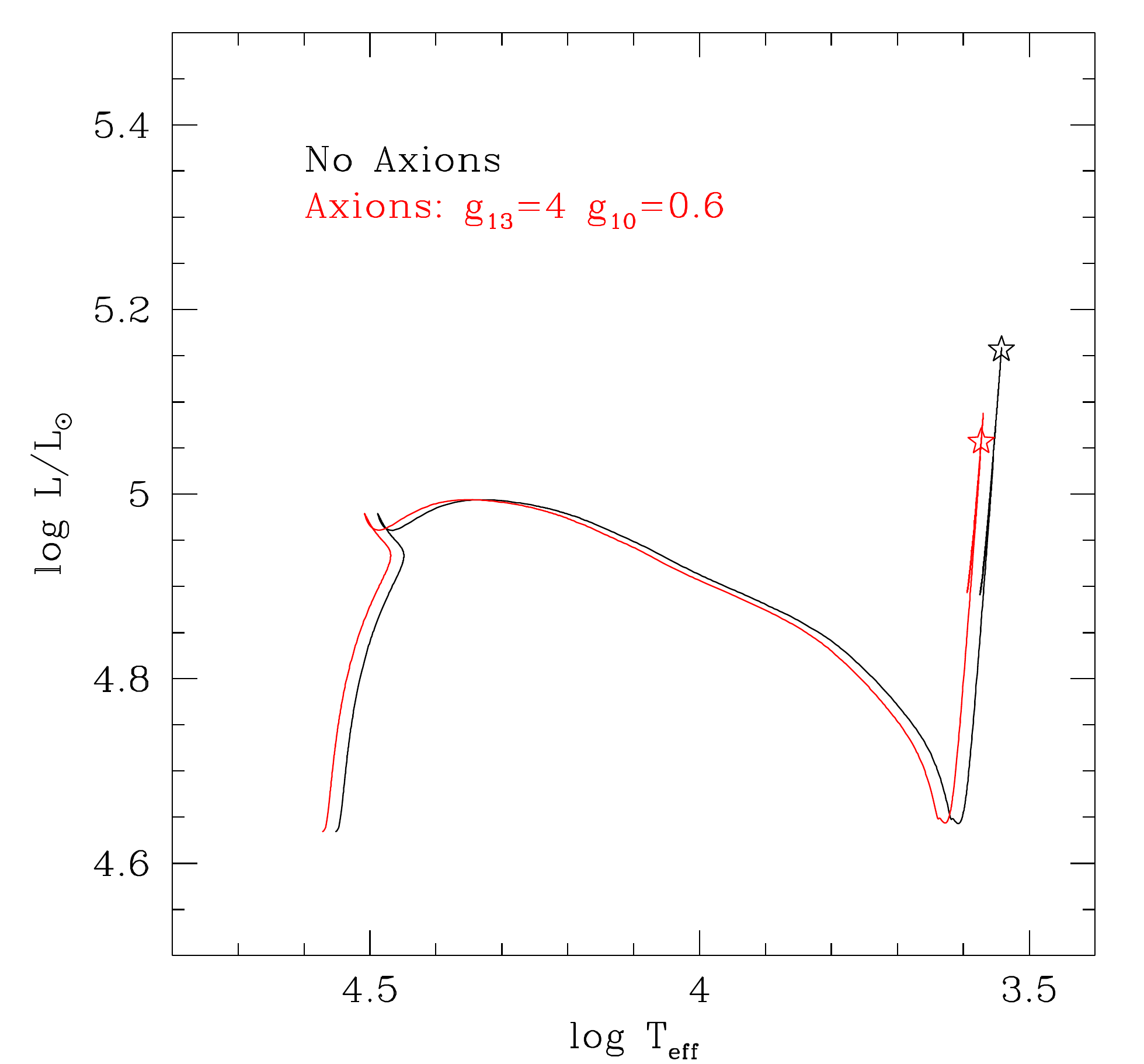}
    \caption{The evolutionary tracks of the two models of Figure \ref{tcroc}. An offset of $\Delta \log T_{eff} = 0.02$ has been applied to the non-standard track. 
   The final $T_{eff},L$ points are marked with a black and a red star, respectively. 
   Note that the model with axions attains a maximum luminosity at the beginning of the C burning.  
   Then, during the C burning, the track moves  to a  slightly fainter point, where remains 
   until the final collapse.}
    \label{cmd20}
\end{figure}

\begin{figure}
    \centering
	\includegraphics[width=14cm,keepaspectratio]{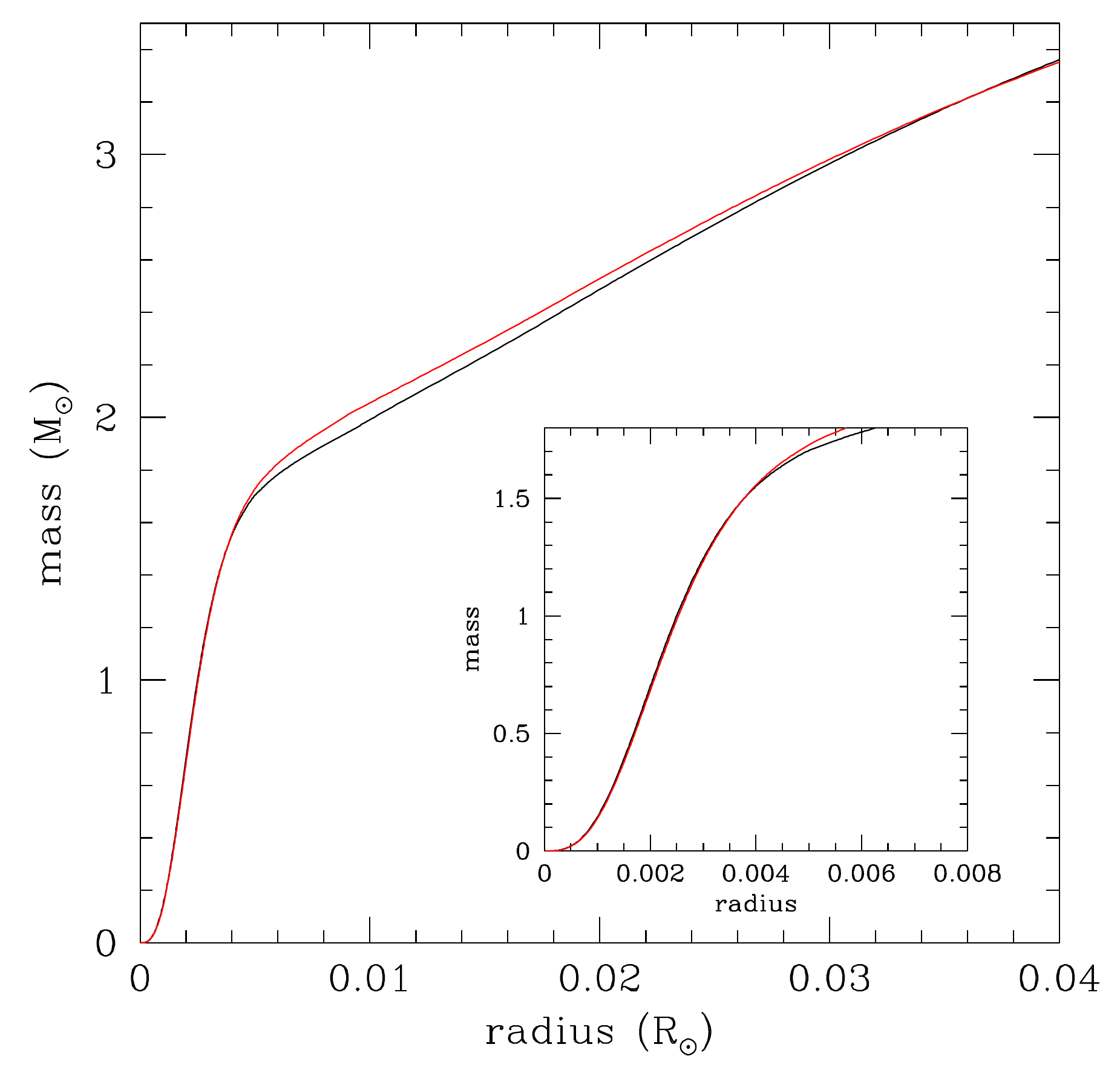}
    \caption{Effects of the axion energy loss on the mass distribution within the core of the last computed models of the two evolutionary tracks shown in Figure \ref{cmd20}. The innermost portion of the core is enlarged in the inset.}
    \label{last}
\end{figure}

The expected axion luminosity along the whole evolution of a 20 M$_\odot$ star is shown in Figure \ref{aloss}. 
 In the right panel, we  show the contributions to the total axion luminosity due to the four production processes described above. 
 It is evident that the Primakoff and the Compton contributions are the major sources of axions in these massive stars. 
 In the left panel, we compare the  total axion luminosity to the thermal neutrino luminosity. 
 Axions clearly represent a significant energy sink in these stars, which is much larger than that induced by neutrinos for a considerable portion of the evolution. 
 In particular, major effects are produced starting from the late part of the He-burning phase. 
 As shown in Figure \ref{tcroc}, when the axion energy loss is considered, the C burning and the Ne burning take place at higher temperature and density. 
 Moreover, the envelope freeze-out occurs earlier than in standard models. 
 Consequently, the final  luminosity is substantially lower (see Figure \ref{cmd20}).
 In addition, axions may also affect the compactness of pre-explosive models. We recall that the evolutionary tracks presented in this paper have been stopped when the maximum temperature attains $\sim 4$ GK. The final mass distribution within the core of the two 20 M$_\odot$ evolutionary tracks, with and without axion energy loss, are compared in Figure \ref{last}. As expected, the axion model shows a higher mass concentration in the 0.006-0.03 R$_\odot$ region.  However, the innermost portion of the core (M$<1.6$ M$_\odot$) is not affected by axion. 
 
\begin{figure}
	\includegraphics[width=\columnwidth]{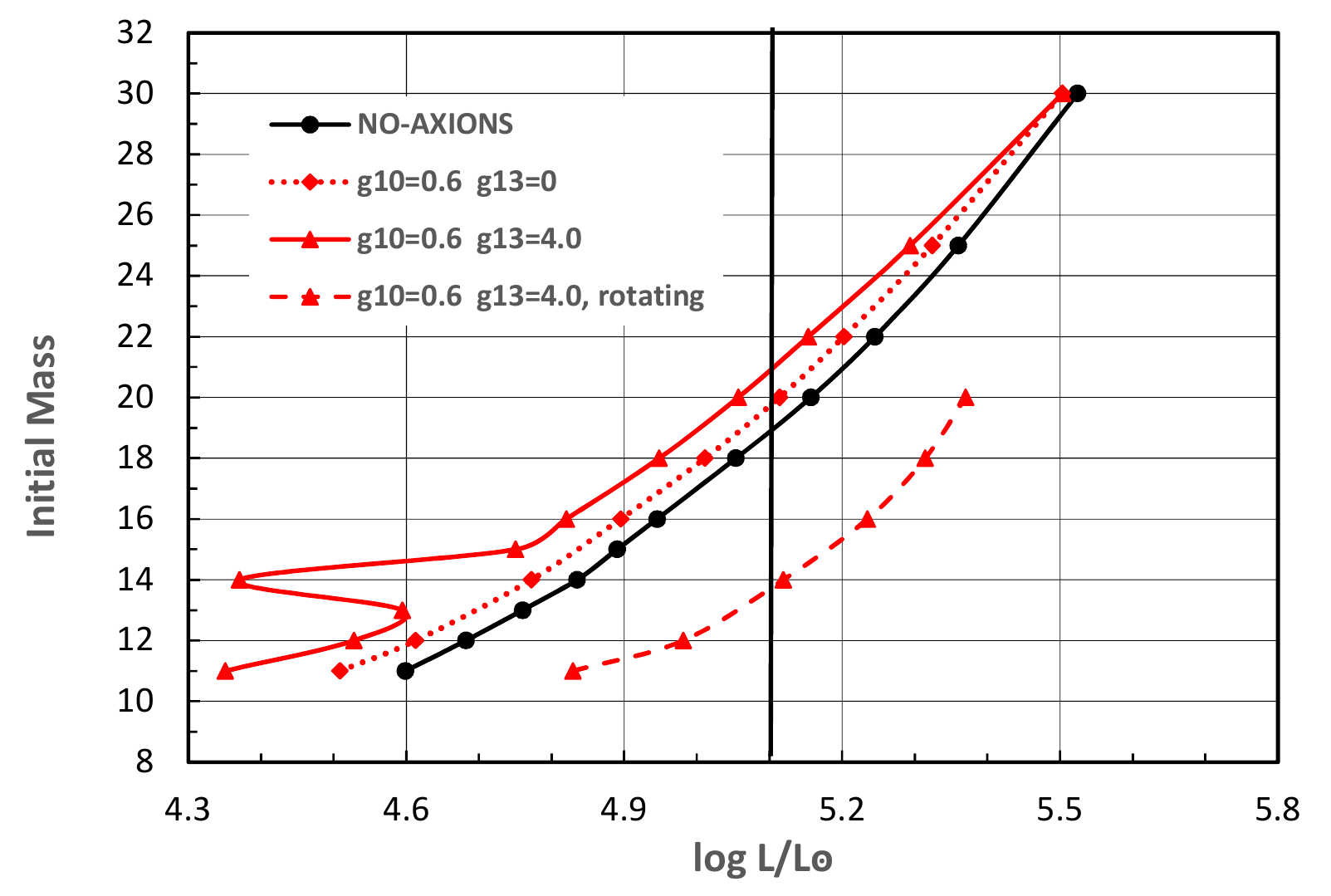}
    \caption{Rotating and non-rotating initial mass-final luminosity relations 
    for models with axion energy sink (red lines)
    are compared to that from non-rotating standard models (black line). 
    The red-dashed line, relative to rotating models, has been obtained 
    with the same set of rotation parameters of the dashed line in Figure \ref{rotation}.}
    \label{axions}
\end{figure}

Then, we have computed two additional set of stellar models. 
In the first one we have included the energy loss induced by the axion-photon coupling alone, 
namely $g_{a\gamma}=0.6\times 10^{-10}$ GeV$^{-1}$ and $g_{ae}=0$.  In this case, only the 
Primakoff process is activated. This choice would be appropriate for, e.g.,  a pure {\it hadronic} axion model, 
such as the KSVZ, or an ALP coupled with photons only, with a coupling close to the astrophysical and experimental upper bound. 
In the second set of stellar models we have switched on the contribution from the 
axion-electron coupling by fixing  $g_{a\gamma}=0.6\times 10^{-10}$ GeV$^{-1}$ and 
$g_{ae}=4\times 10^{-13}$, corresponding roughly to the limit from globular cluster stars. 
The initial mass-final luminosity relations for non-rotating and rotating massive star models 
are reported in Figure \ref{axions}.

The initial masses corresponding to a final luminosity $\log L/L_\odot = 5.1$ are 19.7 and 20.9 
for the models with the axion-photon coupling only and the models with both the axion-photon 
and the axion-electron couplings, respectively.

\begin{figure}
     \centering
	\includegraphics[width=14cm,keepaspectratio]{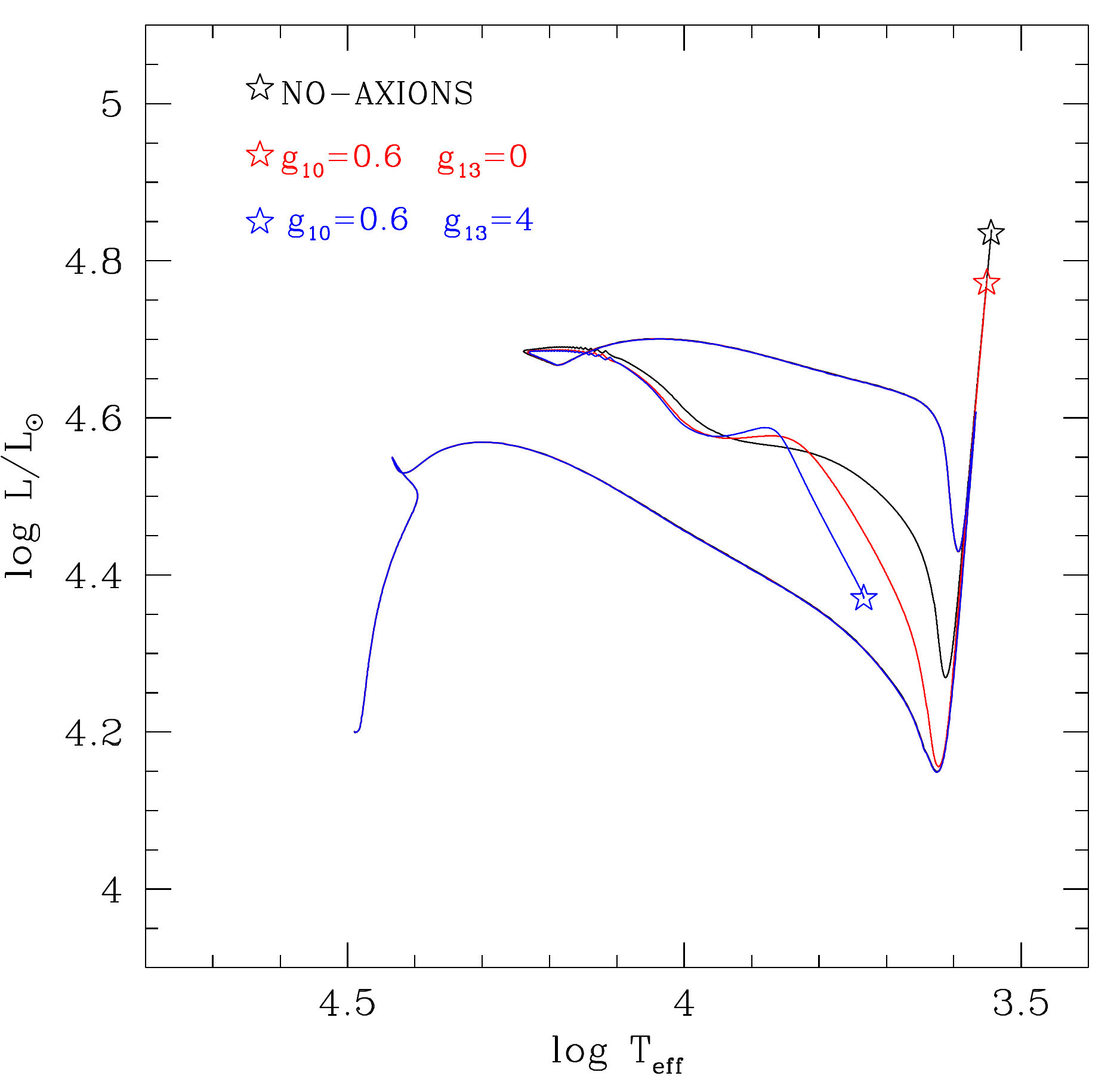}
    \caption{Evolutionary tracks of the 14 M$_\odot$ models with and without axions.}
    \label{cmd14}
\end{figure}

Worth of notice is the peculiar low final luminosity we found in the case of the 14 M$_\odot$ model, namely $\log L/L_\odot = 4.37$, of the second set of axion models. The plot of the corresponding evolutionary track illustrates the origin of this result (see Figure \ref{cmd14}). Indeed this model experiences an extended blue loop during the He burning phase, but at variance with the standard evolutionary track (no axions), the model with both the axion-photon and axion-electron couplings does not complete this loop returning on the red supergiant branch at the end of the He burning. 
Instead the luminosity freeze-out occurs before the star closes its blue loop, 
so that the final point is substantially fainter and bluer than that obtained in the standard case. 
We found that this behaviour affects the evolutionary  tracks of models with mass in a 
restricted range around the 14 M$_\odot$, while the 13 M$_\odot$ model, 
that also experiences a blue loop, shows a normal behaviour. 

\subsection{A new hint of axions? Experimental considerations}
We have shown that axions would substantially modify the initial mass-final luminosity relation 
of massive stars, thus moving to higher values the mass range of type II SN progenitors. 
Axion also represent an appealing solution for various stellar anomalies, 
such as those related to the white dwarf cooling rate.
The considerable larger amount of white dwarfs and type II progenitors expected from the LSST 
deep photometric survey 
will definitely shed light on these problems~\citep{Drlica-Wagner:2019xan,Bechtol:2019acd}.  
Moreover, the axion experimental proposals of the next generation could possibly test this hypothesis. 
In order to make our results more clear from an experimental point of view and to guide 
the experimental effort, here we make some general considerations and 
study the potential of proposed axion experiments. 

As seen, the emission of axions shifts up the initial mass that corresponds to $\log L/L_\odot=5.1$, where $ L $ is the final stellar luminosity.
The exact shift depends on several physical parameters. 
However, we have found a reasonable parametrization as the sum of terms  proportional to the square of the coupling constants:\footnote{The relation is only an approximate fit. 
	A more accurate relation would require the coefficients to depend on $ \Delta M $.
	However, the result is fairly accurate for $ \Delta M \approx 0.5-3 M_{\odot} $.
}
\begin{equation}
\frac{\Delta M}{M_\odot}\approx 2.2\left(\frac{g_{a\gamma}}
{10^{-10}}\right)^{2}+\frac{1.5}{16}\left(\frac{g_{ae}}{10^{-13}}\right)^{2}
\end{equation}
\label{eq:mass_shift}

\begin{figure}
    \centering
	\includegraphics[width=14cm,keepaspectratio]{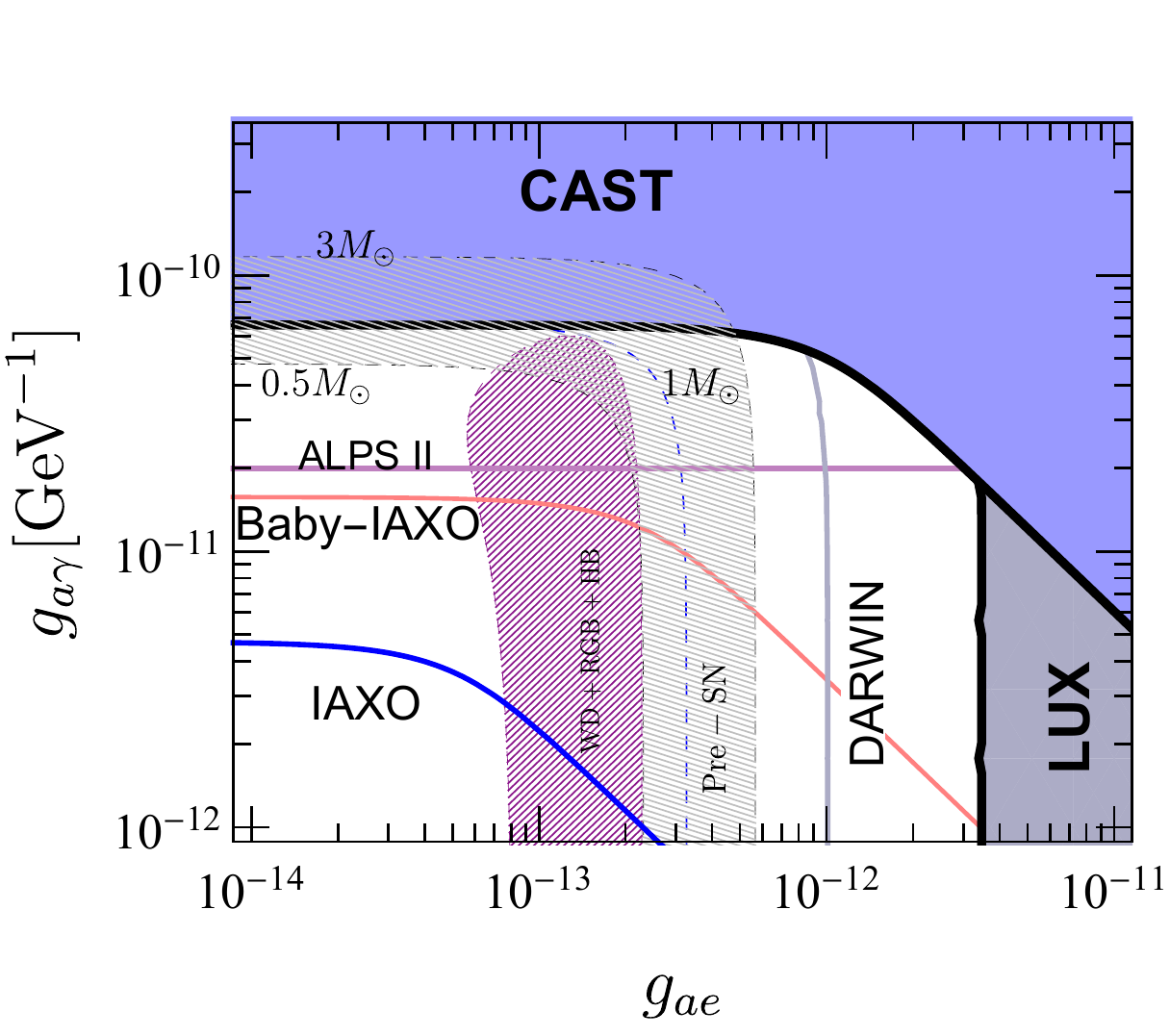}
	\caption{Mass shift for generic ALPs in comparison with the current and planned experimental potential. The parameter regions covered in color are already excluded by current experiments.
		The expected experimental potential of the next generation of axion experiments is indicated with continuous lines. 
		The hushed purple region indicates the 2-$ \sigma $ axion hints from low mass stars~\citep{Giannotti:2016drd,DiVecchia:2019ejf} while the gray hushed region covers the axion parameters that would give $ \Delta M $ between 0.5 and 3$ M_{\odot} $.
	The line corresponding to $ \Delta M =1M_{\odot}$ is also indicated, for reference. 
	}
	\label{fig:sensitivity}
\end{figure}

The parameter band corresponding to the mass shift $ 0.5 M_{\odot} \leq \Delta M\leq 3 M_{\odot}$, as derived from the above equation,  is shown as a gray hushed area in Figure \ref{fig:sensitivity}.
The Figure shows also the experimental potential to explore the region with proposed experiments including BabyIAXO and IAXO [citation], ALPS II~\citep{Bahre:2013ywa} and DARWIN~\citep{Aalbers:2016jon}.
In the case of CAST, BabyIAXO, IAXO and ALPS II, the experimental potential assumes low mass ALPS. 
The region hushed in purple in the Figure is the one inferred from observations of low mass stars~\citep{Giannotti:2016drd}.
The area overlaps, albeit in a small region, with the parameters required for a mass shift $ \Delta M\lesssim 1 M_{\odot} $.
Larger mass shifts require a larger electron coupling (larger photon couplings are already excluded by CAST, at least at mass below $ \sim $ 20 meV).
In any case, the parameter range is accessible to the next generation of axion helioscope experiments as well as to ALPS II. 
On the other hand, as evident from the Figure, WIMP detectors such as LUX, or even the future DARWIN detector, do not have the capability to probe ALP parameters that give a mass shift up to a few $ M_{\odot} $.  

\begin{figure}
 \centering
	\includegraphics[width=14cm,keepaspectratio]{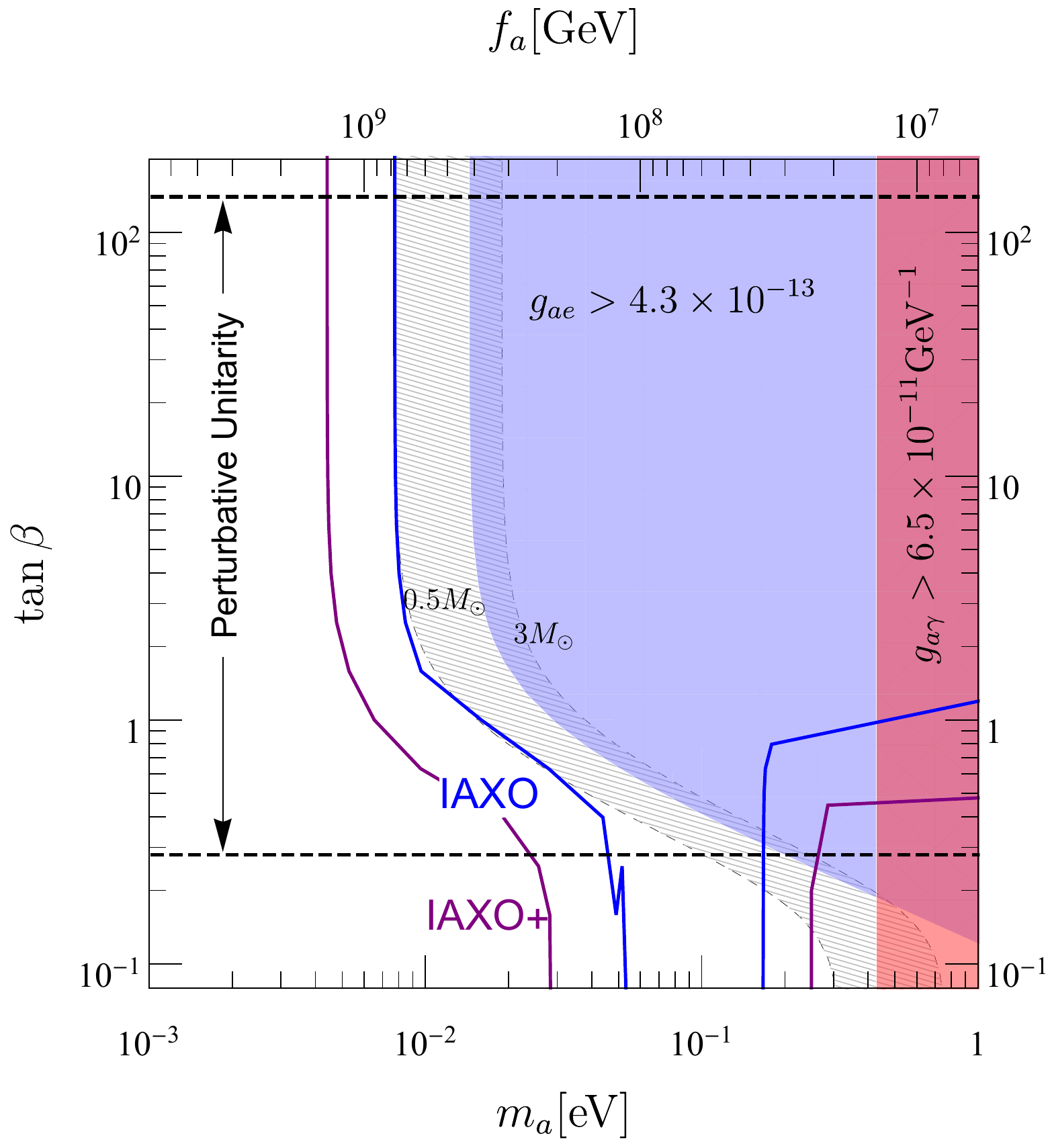}
	\caption{Axion parameter space for the DFSZ axion model, with the expected IAXO experimental potential~\citep{Armengaud:2019uso} and 
		the regions excluded by other astrophysical considerations: $ g_{a\gamma}>6.5 \times 10^{-11}$ GeV$ ^{-1} $ (light red region), from the R-parameter~\citep{ayala2014}, and $ g_{ae}> 4.3\times 10^{-13}$ (light blue region), from the luminosity of tip of the RGB stars in M5~\citep{viaux2013}.
		The set of axion parameters that give a progenitor mass shift between $ 0.5 M_{\odot} $ and $ 3 M_{\odot} $ is enclosed in the gray hushed region. 
		The RGB bound corresponds to $ \Delta M\approx 2 M_{\odot} $.
		The regions of $ \tan \beta $ below 0.28 and above 140 are excluded by the requirement that the Yukawa couplings to fermions satisfy perturbative unitarity.
		The bound from CAST, BabyIAXO and ALPS II do not apply in this case.
	}
	\label{fig:DFSZ_axion}
\end{figure}

QCD axions deserve a separate analysis. 
Because of the mass-coupling relation, the axion couplings of interest for stellar evolution imply masses of $ \sim  $10-100 meV, and could not be detected by the helioscope experiments, except for IAXO. 
Figure \ref{fig:DFSZ_axion} shows the parameter space for the DFSZ axion, which may interact with both electrons and photons, together with the regions excluded by stellar evolution considerations.
The main result is that an upward shift of about $ 2 M_{\odot} $ of the initial masses of SN IIP progenitor 
is allowed when the RGB and HB bounds to the axion couplings are assumed. 
Moreover, IAXO would definitely be able to test the possibility that DFSZ axions are indeed responsible for the missing energy required in our problem.


\section{Summary and Conclusions}\label{sec:discussion}
In this paper we have revised the initial mass-final luminosity relation  
obtained from state-of-the-art massive star models. Standard and
 non-standard physical processes which determine 
this relation are discussed in some details. We find, in particular, 
that different combinations of initial
parameters, such as the mass and the rotational velocity, 
lead to similar final-He-core masses and, in turn, to similar pre-explosive luminosities. 
Moreover, the final luminosity also depends on the initial composition. 
In other words, the final luminosity is not a 
function of only one variable, i.e., the initial mass.
This degeneracy of the initial parameters is often ignored when the 
pre-explosive luminosity is used to estimate the progenitor mass.
In the rest of this section we will analyze this issue.

\begin{figure}
    \centering
	\includegraphics[width=14cm,keepaspectratio]{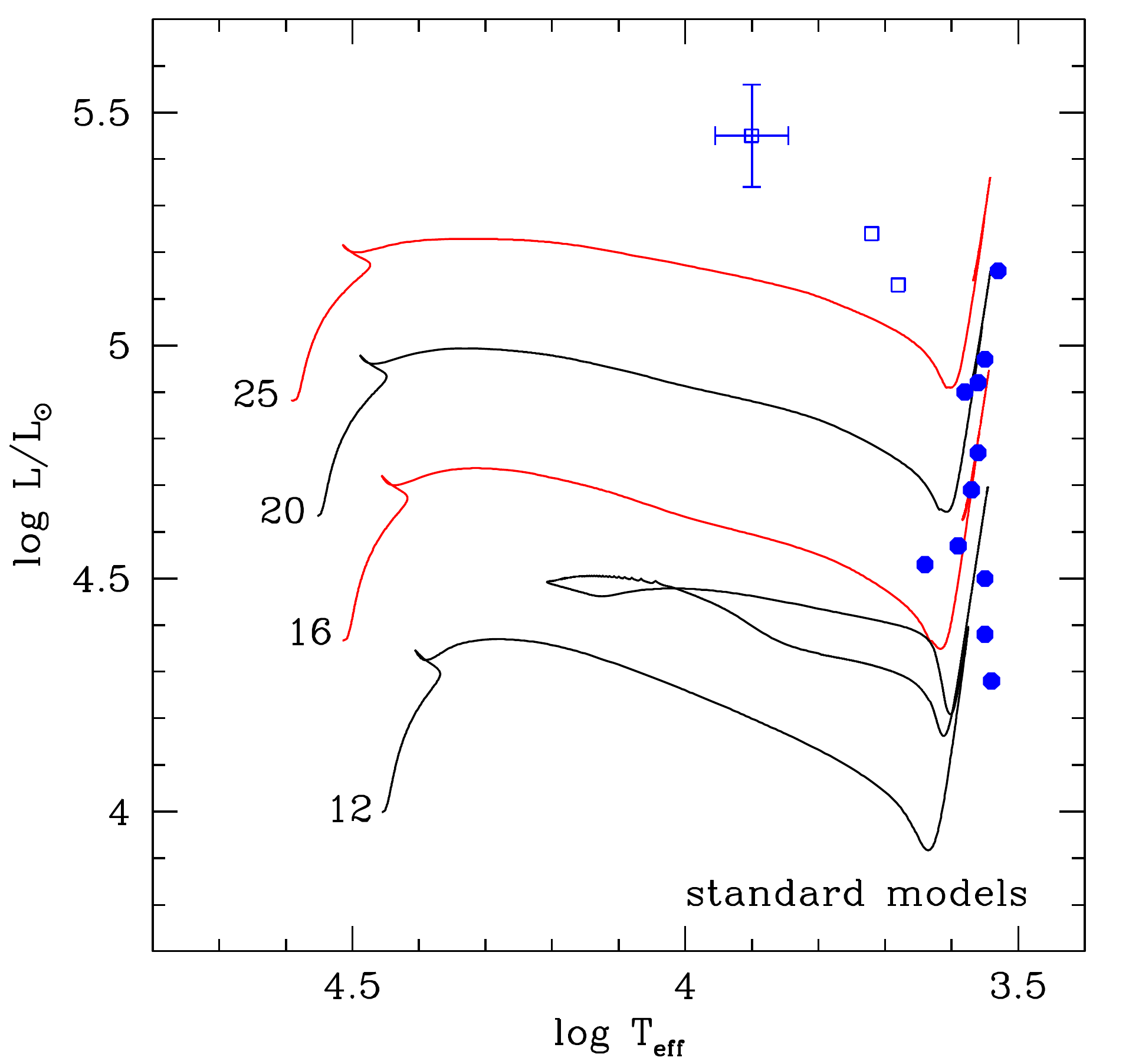}
    \caption{Standard evolutionary tracks (no rotation, no axions).
    The data-points represent SN IIP progenitors (filled circles) and SN IIL 
    progenitors (open squares), 
    from \citet{davies2018}. 
    The average error bar is shown.}
    \label{tracce_std}
\end{figure}

In the HR diagram reported in Figure \ref{tracce_std}, the pre-explosive $\log T_e,\log L$ positions
 of type II supernovae progenitors \citep{smartt2015,davies2018} are superimposed to some 
 of our standard (non-rotating) evolutionary tracks. The two open square refers to type IIL supernovae,
 while the filled circle represent type IIP supernovae.
 Notice that after the revision of the bolometric corrections discussed in \citet{davies2018},
 the upper bound of the pre-explosive luminosity is now slightly larger than 
 that  previously found by \citet{smartt2015}. 
 According to \citet{davies2018}, three progenitors 
 show pre-explosive luminosity $\log L/L_\odot\sim 5.2$. 
 Worth to notice, two of these three points 
 correspond to the only two type IIL progenitors, SN2009kr and SN2009hd,
 while the third, SN2012ec, we suspect it was a fast rotating main-sequence stars (see below). 
 The initial masses estimated 
 by means of the FuNS evolutionary tracks ($v_{ini}=0$) are in the range $9<M/M_\odot <20$, but
 excluding the three brightest objects, the upper bound is just $\sim 17$ M$_\odot$.  

\begin{table}
	\centering
	\caption{Data for SN IIP progenitors. Column 2: pre-explosive $\log L/L_\odot$ from \citet{davies2018}, with
	the corresponding $1\sigma$ error; column 3: most probable mass (in M$_\odot$) 
	from \citep{morozova2018} with the corresponding 95\% C.L. mass range.}
	\label{tabsn}
	\begin{tabular}{lcc} 
		\hline
		 Supernova & $\log L$ &  M  \\
		\hline
 SN2004et &  $4.77 \pm 0.07$ & $16.5_{15.0}^{22.0}$  \\
 SN2005cs &  $4.38 \pm 0.07$ & $ 9.5_{ 9.0}^{12.0}$  \\
 SN2012A  &  $4.57 \pm 0.09$ & $ 9.5_{ 9.0}^{14.0}$  \\
 SN2012aw &  $4.92 \pm 0.12$ & $20.0_{19.0}^{22.5}$  \\
 SN2013ej &  $4.69 \pm 0.07$ & $10.5_{ 9.0}^{18.0}$  \\
 SN2012ec &  $5.16 \pm 0.07$ & $13.0_{10.0}^{18.5}$  \\
 SN1999em &  $<5.02$         & $21.5_{16.5}^{22.0}$  \\     
 SN1999gi &  $<4.85$         & $12.0_{ 9.0}^{17.0}$  \\
  \hline
 \end{tabular}
 \end{table}

\begin{figure}
   \centering
	\includegraphics[width=14cm,keepaspectratio]{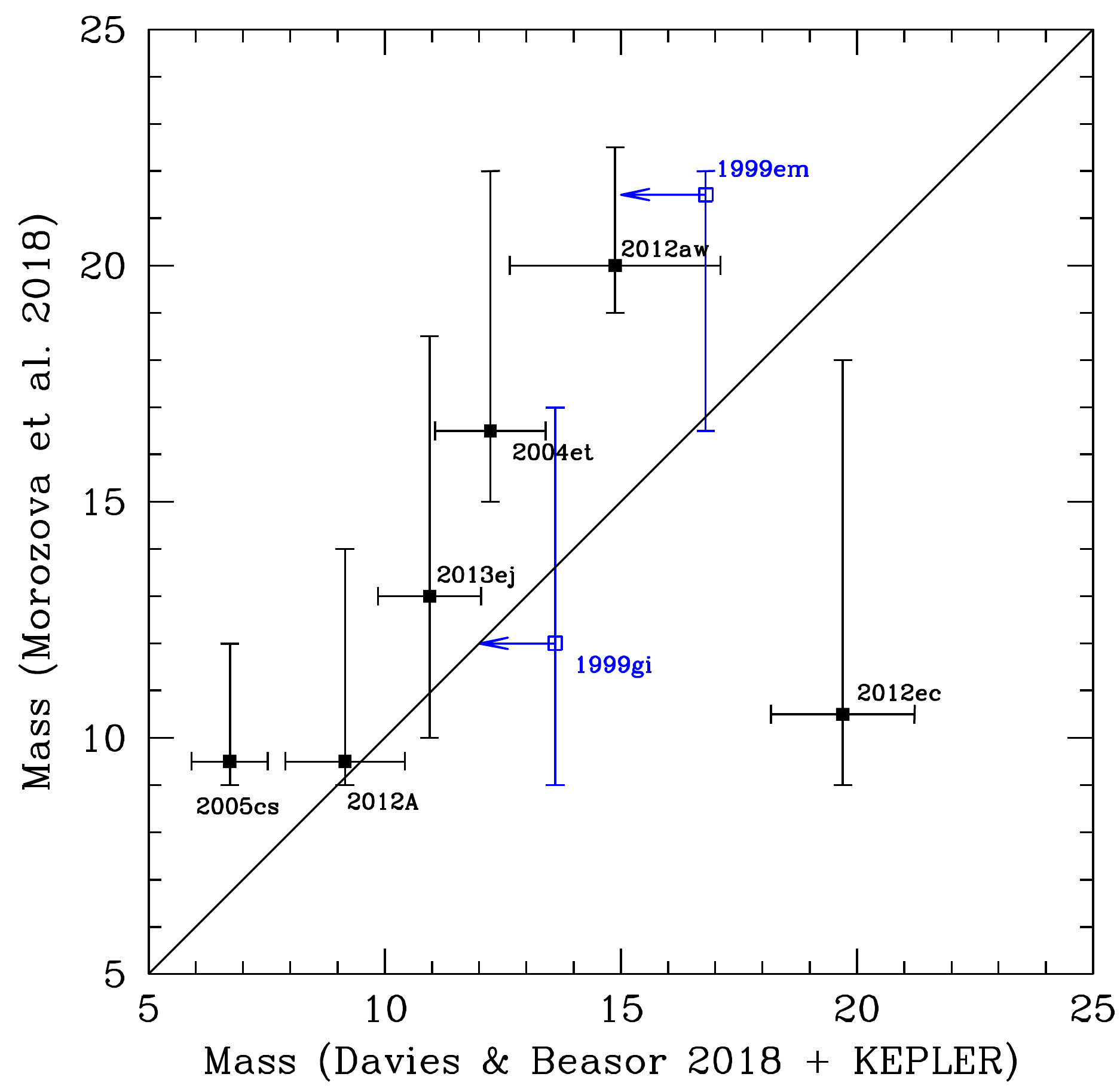}
    \caption{Masses of SN IIP progenitors 
    obtained by \citet{morozova2018} basing on the analysis of the explosive 
    outcomes versus those derived from the observed pre-explosive luminosities 
    \citep{davies2018} and the KEPLER initial mass-final luminosity relation.}
    \label{masse_kepler}
\end{figure} 

\begin{table}
	\centering
	\caption{Masses (in M$_\odot$)
	estimated from the observed pre-explosive luminosities listed in Table \ref{tabsn},
	 as obtained by means of different initial mass-final luminosity relations, namely: 
	non-rotating KEPLER models, non-rotating FuNS models with axion energy loss,
	rotating ($v_{ini}=200$ km s$^{-1}$) FuNS models with axion energy loss. The axion parameters are 
	$g_{10}=0.6$ and $g_{13}=4$. Blank spaces correspond to unrealistic extrapolations 
	of the initial mass-final luminosity relations, i.e., values below 9 M$_\odot$.} 
	\label{tabsn2}
	\begin{tabular}{lccc} 
		\hline 
	    & KEPLER & FuNS+AX & FuNS+AX \\ 
	    & $v_{ini}=0$ & $v_{ini}=0$ & $v_{ini}=200$ \\

		\hline  
 SN2004et & $12.2_{-1.1}^{+1.2}$  & $15.3_{-1.0}^{+1.0}$ &               \\   
 SN2005cs & $                  $  & $11.0_{-0.6}^{+0.5}$ &               \\
 SN2012A  & $ 9.2_{-1.2}^{+1.3}$  & $12.8_{-1.0}^{+1.0}$ &                 \\
 SN2012aw & $14.9_{-2.1}^{+2.3}$  & $17.6_{-1.9}^{+2.1}$ &  $11.4_{-0.9}^{+1.4}$  \\
 SN2013ej & $10.9_{-1.1}^{+1.1}$  & $14.2_{-0.9}^{+0.9}$ &                 \\
 SN2012ec & $19.7_{-1.5}^{+1.5}$  & $22.0_{-1.5}^{+1.4}$ &  $14.8_{-1.2}^{+1.4}$  \\
 SN1999em &  $<16.8$         & $<19.3$         &  $<12.5$             \\     
 SN1999gi &  $<13.6$         & $<16.4$         &  $<10.8$               \\
  \hline
 \end{tabular}
 \end{table}

As mentioned in the introduction an independent estimation of the progenitor mass can be obtained 
from the best fit of the observed properties of SN light curves (LCs). 
Only 8 SNe IIP are in common between the \citet{morozova2018} and the \citet{smartt2015} samples. 
Pre-explosive luminosities and initial masses
of the corresponding progenitors are listed in Table \ref{tabsn}.
Although independent on the  
final luminosity, the LC best fit  method also relays on theoretical models. 
\citet{morozova2018} have used, in particular, KEPLER progenitors to model the outgoing shock
and its interaction  with the dense circumstellar material surrounding the exploding progenitor.  
Therefore, 
in Figure \ref{masse_kepler} we compare their results with the masses derived from the pre-explosive 
luminosity and the initial mass-final luminosity relation from (non-rotating)
 KEPLER models. Nonetheless, negligible differences of this Figure would have been found
  if we had used our initial mass-final luminosity relation.
 
 Six out of eight data points clearly stay above the median, while just one, SN2012ec, 
 is definitely below it. Excluding SN2012ec, 
 it results that the masses from the LC best fit are $2.91\pm 0.84$ M$_\odot$ higher, 
 on the average, than those 
 estimated from the pre-explosive luminosity. 
 This discrepancy worsens when convective-core 
 overshoot or rotation are taken into account (see section \ref{sec_standard}). Indeed, 
 both these processes imply larger He-core masses and, in turn, brighter progenitors, so that the masses derived from the corresponding 
 initial mass-final luminosity relation are smaller. 
 Other missing physical processes, like enhanced mass loss due to a binary mass transfer episode (section \ref{binary}) or envelope  heating driven by internal gravity waves (section \ref{massloss}),  would also worsen the tension.
 
If the effect of convective-core overshoot may be negligible, rotation is a quite common feature 
of massive stars. According to \citet{hunter2008,hunter2009}, 
 the projected rotational velocity ($v\sin i$) of 
  massive H-burning stars ranges between 0 and 300 km s$^{-1}$ and, when the mass range is limited to 
  stars with $M<25$ M$_\odot$, the distribution peaks between 100 and 150 km s$^{-1}$. This evidence
  reinforces our claim of a discrepancy between the masses
estimated by means of the SN light curve and those derived from   
pre-explosive luminosities. Indeed,  
when rotational velocities of this order of magnitude are considered, 
the masses estimated from the final luminosity 
are up to 6 M$_\odot$ smaller (see section \ref{sec_rotation} and Table \ref{tab1}). Fast rotation may be eventually invoked to explain SN points located below the median in Figure 
\ref{masse_kepler}. For instance, even the extreme case of SN2012ec can be fitted by assuming 
a large initial rotational velocity ($\sim 200-300$ km s$^{-1}$). Indeed, the mass inferred from the LC best fit is significantly smaller than that derived from 
the pre-explosive luminosity of non-rotating models \citep{barbarino2015,morozova2018}. This occurrence is also confirmed by the mass estimate independently obtained from the O yield of the SN2012ec \citep{valenti2016,davies2018}. 
A spread of 
initial rotational velocities also provide a natural explanation 
of the lack of a direct correlation between initial mass and final luminosity.

\begin{figure} 
    \centering
	\includegraphics[width=14cm,keepaspectratio]{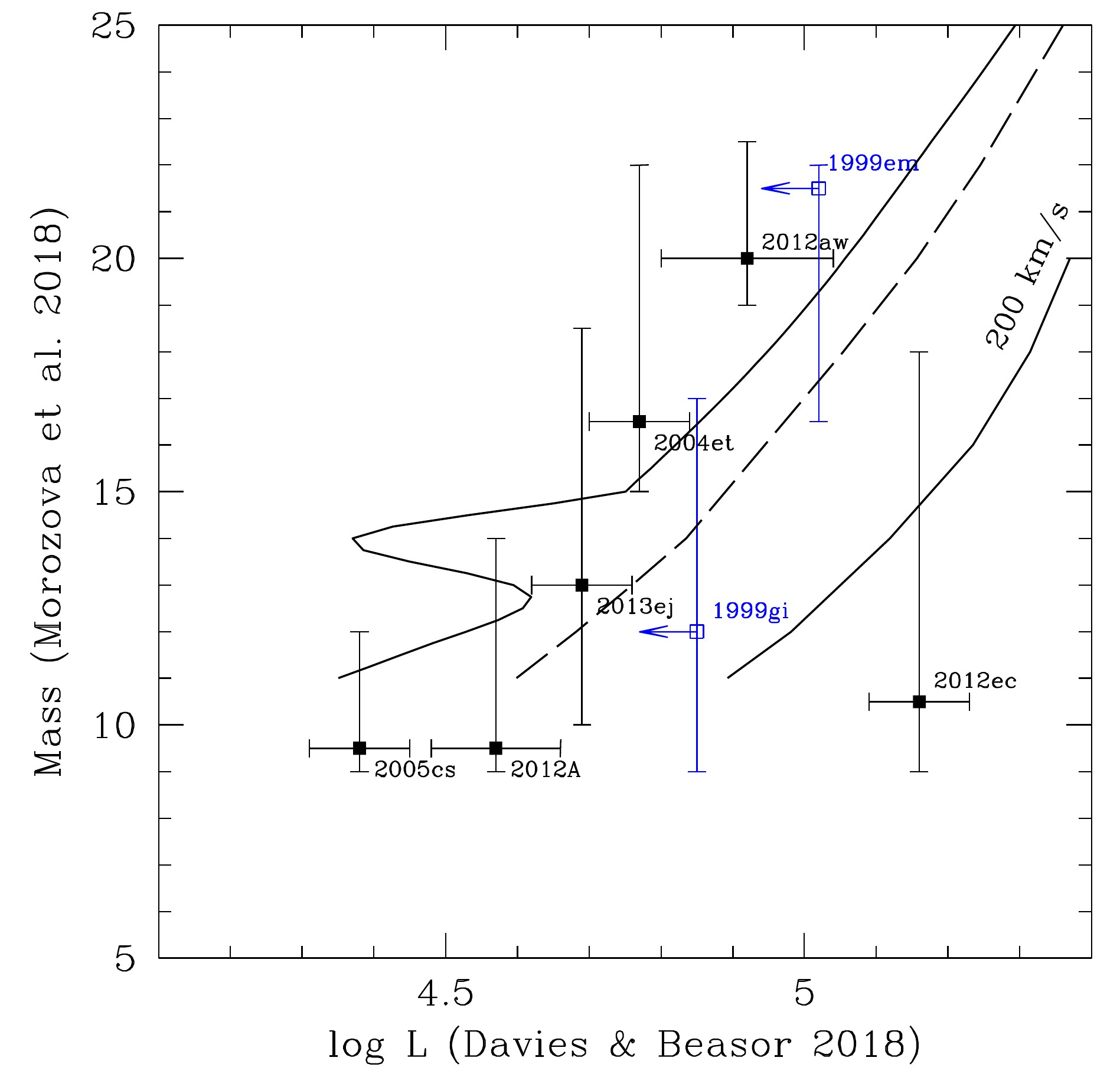}
    \caption{Theoretical versus observed initial mass-final luminosity relation. The observed mass and luminosity are those 
    listed in Table \ref{tabsn}. The theoretical relations (black-solid lines) are derived from 
    FuNS models with axion energy loss 
    ($g_{10}=0.6$, $g_{13}=4$) and with or without rotation. 
    For comparison, the FuNS standard relation is also shown (dashed line).}  
    \label{mlrelax}
\end{figure}

In any case, the initial mass-final luminosity relation 
from non-rotating models should represent the upper bound to the initial masses. 
In contrast, some progenitors, such as SN1999em, SN2004et and SN2012aw, show much higher masses 
than expected.  For example, in the case of SN2004et, the maximum 
mass inferred from the pre-explosive  
luminosity is $12.2\pm{1.1}$ M$_\odot$, while Morozova et al. found a most probable mass of 
16.5 M$_\odot$, with 
a 95\% C.L range between 15 and 22 M$_\odot$. Notice that a previous estimate of the mass range
for this supernova progenitor based on LC best fit  
was 24-29 M$_\odot$ \citep{utrobin2009}. Similarly, for SN2012aw Morozova et al. found 
a most probable value
of 20 M$_\odot$ (19-22 M$_\odot$ is the 95\% C.L. range, see also \citet{barbarino2015}), 
while the maximum mass 
compatible with the observed pre-explosive luminosity, as obtained from non-rotating 
(no-overshoot) models, is just $14.9\pm 2.3$ M$_\odot$ (see Table \ref{tabsn2}). 

In spite of the limited sample of supernovae, our analysis suggests that behind 
this problem there is a clue of missing physical processes.
The most natural way to reduce 
the final luminosity and, in turn, allow more massive progenitors, 
is to increase the rate of energy loss. 
As seen in the previous section, 
the inclusion of axion production in the calculations of massive star models 
would imply a sizable increase of the progenitor masses estimated by means of the 
observed pre-explosive luminosity.
In particular, depending on the assumed couplings of axions with other standard particles 
(cfr. Figure \ref{fig:sensitivity} and~\ref{fig:DFSZ_axion}), 
the resulting progenitor masses are up to 
$ \sim 3M_{\odot} $ larger than those obtained without axions. As an example, in Table \ref{tabsn2} we report the masses of the 8 SNe progenitors already listed  in Table \ref{tabsn} as obtained by means 
of the initial mass-final luminosity relations derived from models with axion, by assuming
$g_{a\gamma}=0.6\times 10^{-10}$ GeV$^{-1}$ and $g_{ae}=4\times 10^{-13}$, 
for the axion-photon and the axion-electron 
couplings, respectively.
  The axion scenario is  illustrated 
in Figure \ref{mlrelax}, where we show the initial masses from the LC best fit  \citep{morozova2018}
versus the observed pre-explosive luminosity \citep{davies2018}, 
but superimposed are the theoretical relations we obtained including axions 
and initial rotational velocity $v_{ini}=0$ and 200 km s$^{-1}$. 
For comparison, we also show the standard relation (no-axion) for non-rotating models.

Summarizing, we have shown that the initial mass-final luminosity relation is a powerful tool 
to investigate physics beyond the Standard Model. 
In spite of the limited data sample, the available measurements appear 
incompatible with the standard picture and suggest 
possible hints in favor of an enhanced energy loss rate from weakly interactive particles. 
In particular, once a spread of initial rotational velocity is assumed in agreement with 
the extant measurements of this quantity, the best reproduction of the available sample of 
initial mass-final luminosity data is obtained with models that account for the production of 
axions or ALPs.   
The choice of axions/ALPs as the new physics candidates proposed to solve this problem 
is dictated by the current interest in axion research and the confidence that our hypothesis may be tested in the near future. 
However, we cannot exclude that other new physics processes are at work. For instance a similar effect could be caused by the thermal production and emission of dark photons. On the base of the  present analysis, such a possibility is not excluded.

\acknowledgments 
O.S and L.P. are supported by the Italian Space Agency (ASI) and the 
Italian National Institute of Astrophysics (INAF) under the agreement n. 2017-14-H.0 -
 {\it attivit\'a di studio per 
la comunit\'a scientifica di Astrofisica delle Alte Energie e Fisica Astroparticellare}.
I.D. is supported by the MICINN-FEDER project AYA2015-63588-P and A.M. by the
Istituto Nazionale di Fisica Nucleare (INFN) through the ``Theoretical Astroparticle Physics'' 
project and by Ministero dell'Istruzione, Universit\`a e Ricerca (MIUR).

\appendix

\section{Axion emission rates}
\label{sec:appendix}
In this appendix we provide the numerical recipes used to calculate the axion emission rates.
The density of the $j$ ions is 
\begin{equation}
 n_j =\frac{\rho}{m_u}\frac{X_j}{A_j}\,,
\end{equation}

where $m_u =1.66 \times 10^{-24} {\rm g} $ is the 
atomic mass unit and $X_j$, $A_j$, and $Z_j$ are the mass fraction, 
the atomic mass, and the charge of the $j$ specie, respectively. 
Then, the electron density is
\begin{equation}
 n_e =\frac{\rho}{m_u\mu_e}\,,
\end{equation}
 
with the mean electron molecular weight defined by
\begin{equation}
 \frac{1}{\mu_e}=\sum\frac{X_j Z_j}{A_j}\,,
\end{equation}

In the following, the temperature, $T$, and the mass density, $\rho$, are in K 
and $ {\rm g}\,{\rm cm}^{-3} $, respectively.
We indicate with $\varepsilon$ the energy loss rates in erg g$^{-1}$ s$^{-1}$.

\subsection{Primakoff production}
\label{sec:Primakoff}
The Primakoff production of axions in a stellar core, that is the production of axions from thermal photons conversion in the electrostatic field of the nuclei or of the electrons
$\gamma + Ze \to \gamma +a $,
is widely discussed in the literature (e.g., \citet{Raffelt:1985nk,Raffelt:1987yu}). 
Our method is based on \citet{Raffelt:1987yu}, but includes a new parametrization of the
 degeneracy effects.  

The axion emission rate is calculated summing the contributions from the scattering of 
photons on electrons and ions.
We assume that ions are non-degenerate while
electrons may have a certain degree of degeneracy. 
The Coulomb interaction between charges and thermal photons is screened.
The typical screening scale is related to the longitudinal component of the polarization tensor which, 
in the static and non-relativistic limit is \citep{raffelt_book}
\begin{equation}
\pi_L=\frac{4 Z^2\alpha\, m}{\pi}\int_{0}^{\infty} \frac{1}{\exp[\beta(E-\mu)]+1}\,dp\,,
\end{equation}
where $ \beta=1/k_B T $, with $ k_B $ the Boltzmann constant, $ \mu $ is the chemical
potential and $ E = m + p^2/2m $ is the fermion energy.
In the limit of a non-degenerate plasma, $ \pi_L $ converges to 
\begin{equation}
	\pi_L\to \kappa_D^2=\frac{4\pi Z^2\alpha\, n}{T}\,,
\end{equation}
where $ n $ is the number density of the charged particles,
setting the Debye length as the typical screening scale. 
For a very degenerate plasma, on the other hand, the relevant screening parameter becomes the Thomas-Fermi scale
\begin{equation}
\pi_L\to \kappa_{TF}^2=\frac{4 Z^2\alpha\, m\,p_F}{\pi}\,.
\end{equation}
In general, to account for any degree of degeneracy it is convenient to introduce the degeneracy parameter $ R_{\rm deg} $, defined such that
\begin{equation}
\kappa^2=R_{\rm deg} \kappa_D^2\,.
\end{equation}
We have derived a good numerical fit for the degeneracy parameter as a function of the 
temperature and density of the electron gas. 
Introducing the dimensionless parameter
\begin{equation}
z=\frac{\rho}{T^{3/2}\mu_e}\,,
\end{equation}
where $ T $ is in Kelvin and $ \rho $ in g$ $cm$^{-3} $, we find
\begin{equation}
R_{\rm deg}=1\,,
\end{equation}
for $ z\leq 5.45\times 10^{-11} $;
\begin{equation}
R_{\rm deg}=0.63+0.3\arctan \left( 0.65-9316\,z^{0.48}+\frac{0.019}{z^{0.212}}\right) \,,
\end{equation}
for $ 5.45\times 10^{-11}< z\leq 7.2\times 10^{-8} $;
and 
\begin{equation}
R_{\rm deg}=4.78 \times 10^{-6}z^{-0.667}\,,
\end{equation}
for $ z> 7.2\times 10^{-8}$.
The precision of the fit is always better than 10\%, in the whole region of interest.\footnote{Notice that this numerical result can be applied only to electrons, not to protons since we used explicitly the electron mass.}

A further effect of the partial degeneracy is the reduction of the effective number of targets $ n\to n_{\rm eff} $, defined, for example, in \citet{Raffelt:1987yu}.
As shown in \citet{Payez:2014xsa}, this correction factor is again $ R_{\rm deg}$. 
Therefore, we find $n_{\rm eff} =R_{\rm deg}\, n$.

We are now ready to provide a numerical recipe for the Primakoff axion production rate.
Let us first define the quantities
\begin{equation}
y_{pl}=\frac{\omega_{\rm pl}}{T}=\frac{\left(\rho/\mu_e\right)^{1/2}}{\left( 1+ \left( 1.02 \times 10^{-6}\rho/\mu_e\right)^{2/3}\right)^{1/4}}\,,\nonumber \\ 
\end{equation}

\begin{equation}
y_{\rm ions}=2.57\times 10^{10}\left(\frac{\rho}{T^3}\sum_{\rm ions} \frac{Z_j^2 X_j}{A_j}\right)^{1/2}\,,\nonumber \\
\end{equation}

\begin{equation}
y_{\rm el}=2.57\times 10^{10}R_{\rm deg} \left(\frac{\rho}{T^3}\sum_{\rm ions} \frac{Z_j X_j}{A_j}\right)^{1/2}\,,\nonumber \\ 
\end{equation}

\begin{equation}
y_s=\left( y_{\rm ions}^2+y_{\rm el}^2\right)^{1/2}
\end{equation}
where $ T $ is in K and $ \rho $ in $ {\rm g}\,{\rm cm}^{-3} $, and the function
\begin{eqnarray}
f(y_{pl},y_s)=\frac{1}{4\pi}\int_{y_{pl}}^{\infty} \frac{y^2 \sqrt{y^2-y_{pl}^2}}{e^{y}-1}\,I(y_{pl},y_s)\, dy\,.
\end{eqnarray}
with 
\begin{equation}
I=
\frac{r^2-1}{s}\ln\left( \frac{r-1}{r+1}\right)+
\frac{(r+s)^2-1}{s}\ln\left( \frac{s+r+1}{s+r-1}\right) -2\,, 
\end{equation}%
and 
\begin{equation}
r=\frac{2y^2-y_{pl}^2}{2y\sqrt{y^2-y_{pl}^2}}\,, \qquad
s=\frac{y_s^2}{2y\sqrt{y^2-y_{pl}^2}}\,. 
\end{equation}%
Therefore, the axion Primakoff emission rate reads
\begin{equation}
\varepsilon=4.71\times 10^{-31}g_{10}^2T^4
\left[\sum_{\rm ions} \left(Z_j^2+R_{\rm deg} Z_j\right)  \frac{X_j}{A_j}\right] f(y_{pl},y_{s}) 
\frac{\rm erg}{{\rm g}\cdot {\rm s}}\,,
\end{equation}
with $ g_{10}=g_{a\gamma}/10^{-10} $GeV$ ^{-1} $ and $ T $ in K.

\subsection{Compton}
\label{sec:Compton}
The Compton axion production, 
$ \gamma +e\to  \gamma +a $, is the production of axions from the scattering 
of thermal photons on electrons and it is driven by the axion-electron coupling.  
The non-relativistic cross section is \citep{raffelt_book}
\begin{equation}
\sigma=\frac{1}{3}\alpha\left(\frac{g_{ae}}{m_e}\right)^2
\left(\frac{\omega}{m_e}\right)^2\,,
\end{equation}
where $ \omega $ is the photon energy, assumed to be $ \omega\ll m_e $. 
Here, we have  neglected the plasma frequency. 
This simplification, however, does  not substantially modify the result.
Additionally, the Compton process is relevant only in the non-degenerate regime since
 in a degenerate plasma Bremsstrahlung dominates. 

The induced energy loss is 
\begin{equation}
 \varepsilon=R_{\rm deg} \frac{n_e}{\rho} \int \frac{2\,d^3 \textbf{k}}{(2\pi)^3}\frac{\sigma\omega}{e^{\omega/T}-1}\,,
\end{equation}
where $ R_{\rm deg} $ accounts for the reduction in number of effective electron targets when the plasma is degenerate.%
\footnote{Notice that the integral can be solved analytically if we neglect $ \omega_{\rm pl} $.
That is what we do in the following, particularly in Eq.~\ref{Eq:Compton}.
Here we are also ignoring the screening.
This is justified since the integral converges (contrarily to what happens in Primakoff and Bremsstrahlung) and so the screening adds only a small correction.}
We already have a fit for $ \theta_{deg} $.
Extracting the dependence on $ \omega $ and setting $ x=\omega/T $ we find
\begin{equation}
 \varepsilon=R_{\rm deg} \frac{n_e}{\rho} \frac{8\pi}{(2\pi)^3}\left( \frac{\sigma}{\omega^2}\right)T^6 \int_{0}^{\infty} \frac{x^5}{e^{x}-1} \,dx \\ \nonumber
 \simeq R_{\rm deg} \frac{122.1}{3\pi^2}\frac{1}{m_e^4\, m_u\,\mu_e} \,\alpha \, g_{ae}^2 \,T^6 \,.
\end{equation}
Numerically:
\begin{equation}
\label{Eq:Compton}
\varepsilon_{\rm compton}=\theta_{deg} 2.66\times 10^{-48}\,g_{13}^2 \,\frac{T^6}{\mu_e}\,
\end{equation}
where $g_{13}=g_{ae}/10^{-13}$.

\subsection{Bremsstrahlung}
\label{sec:Bremsstrahlung}
The Bremsstrahlung process, $ e +Ze\to  e + Ze +a $, is the most important axion production mechanism in a degenerate plasma.
Therefore, we will discuss the degenerate limit first. 
In this limit we neglect the contribution from the scattering on a single electron, since very few electron targets are available in this limit.
Additionally, the Debye screening length is a good approximation for the screening length for ions. 
With these approximations, we find (in units of erg g$^{-1}$ s$^{-1}$)
\begin{equation}
\varepsilon_{\rm BD}=8.6 \,F\times 10^{-33}\,
 g_{13}^2 T^4\,\left(\sum\frac{X_j Z_j^2}{A_j}\right) 
\,,
\end{equation}
where
\begin{equation}
F=\frac{2}{3}\ln\left (\frac{2+\kappa^2}{\kappa^2}\right )+ \left[ \frac{2+5\kappa^2}{15}\ln\left(\frac{2+\kappa^2}{\kappa^2}\right)-\frac{2}{3}\right]\beta_F^2 \,,
\end{equation}
with
\begin{equation}
\kappa^2=\frac{k_D^2}{2p_F^2} =9.27\times 10^{4}\,\frac{\rho^{1/3}}{T} 
\left( \sum\frac{X_j Z_j^2}{A_j} \right)^{1/3}\,, \\
\end{equation}
\begin{equation}
 \beta_F=\frac{p_F}{E_F}=\frac{p_F}{\sqrt{m_e^2+p_F^2}} \,, \\
\end{equation}
\begin{equation}
 p_F=\left( 3\pi^2\frac{\rho}{m_u}\sum\frac{X_j Z_j}{A_j}\right) ^{1/3}=5155 \left( \rho \sum\frac{X_j Z_j}{A_j}\right)^{1/3} {\rm eV}\,.
\end{equation}

In the non-degenerate limit one finds (again, in units of erg g$^{-1}$ s$^{-1}$)
\begin{equation}
\label{Eq:BND}
\varepsilon_{\rm BND}=4.7\times 10^{-25} g_{13}^2 T^{2.5}\,\frac{\rho}{\mu_e}
\sum \frac{X_j\,Z_j}{A_j}
\left(Z_j+\frac{1}{\sqrt{2}} 
  \right) \,.
\end{equation}
Eq.~\ref{Eq:BND} does not take into account the screening effects, which are always small in the limits 
of validity of this expression. To take those effects into account, one has to substitute the sum 
in Eq.~\ref{Eq:BND} with
\begin{equation}\label{Eq:BNDfull}
\sum \frac{X_j}{A_j}
\left[Z_j^2 \left(1-\frac58 \frac{k_S^2}{m_e T}\right) +\frac{Z_j}{\sqrt{2}} \left(1-\frac54 \frac{k_S^2}{m_e T}\right)  \right] 
\end{equation}
where $ k_S $ accounts for the screening which, in the non-degenerate limit, takes contribution from both electrons and ions. Numerically:  
\begin{equation}
	\frac{k_S^2}{m_e T}=1.12\times 10^{11}\,\,\frac{\rho}{T^2}\, \sum\frac{X_j (Z_j+Z_j^2)}{A_j}\,.
\end{equation}

For the intermediate degeneracy case, we use the approach discussed in \citet{Raffelt:1994ry}:
\begin{equation}\label{Eq:RW_fitting}
\varepsilon=\left( \frac{1}{\varepsilon_{\rm D}}+\frac{1}{\varepsilon_{\rm ND}}\right)^{-1}\,.
\end{equation} 

\subsection{Pair Production}
Finally, axions can be produced from the annihilation of an electron-positron pair, $ e^{+} e^{-} \to \gamma + a $. 
According to \citet{Pantziris:1986dc}, we define the dimensionless parameters
\begin{equation}
\lambda=T/m_e\simeq 1.69\times 10^{-10}T_K\,, 
\end{equation}
and
\begin{equation}
\nu=\mu/T\,,
\end{equation}
and distinguish among different plasma conditions. 

In the non-degenerate and non-relativistic regime, we find 
\begin{equation}
\varepsilon=\frac{g_{ae}^{2}\alpha}{4\pi^3\rho} m_e^2 T^3 e^{-2/\lambda}\left( 1+\frac{\lambda}{2}\right)\,.
\end{equation}

The degenerate case is given in \citet{Pantziris:1986dc} and applies to the case of $ \lambda\ll 1 $ and $ 1/\lambda\ll \nu \ll 2/\lambda $.
The result is cumbersome 
\begin{equation}
\varepsilon=\frac{g_{ae}^{2}\alpha}{\pi^4\rho} m_e^2 T^3 e^{\eta}
\int_{0}^{\infty}\frac{dx\sqrt{x}}{e^{x-a}+1}
\int_{0}^{\infty}\frac{dy\sqrt{y}}{e^{y} }\left( 1+\frac{x+y}{6}\lambda\right) \,,
\end{equation}
where $ \eta=\nu-1/\lambda $.\footnote{Notice that, in the limit of validity of this approximation, $ \eta<1/\lambda $.}
A simplified expression, accurate to about 15\% in the whole range of interest, is  
\begin{equation}
\varepsilon=\frac{g_{ae}^{2}\alpha}{\pi^4\rho} m_e^2 T^3 e^{\nu-1/\lambda}f(\nu-1/\lambda)\,,
\end{equation}
where
\begin{equation}
f(x)=0.605\, e^{-0.84 x}\,.
\end{equation}

Finally, in the non-relativistic and mildly degenerate case ($ \mu\simeq m_e$), we find the expression 
\begin{equation}
\varepsilon=\frac{g_{ae}^{2}\alpha}{4\pi^3\rho} m_e^2 T^3 e^{-2/\lambda}0.76\left( 1+0.53\lambda\right) \,.
\end{equation}

A numerical expression for the axion pair production in cgs units
that summarizes all the previous results is
\begin{equation}
\varepsilon =\frac{2.14\times 10^{16} g_{13}^2 \lambda^3}{\rho }\,\frac{{\rm erg}}{{\rm g \, s}} \, e^{-{\rm Max(\eta,0)}-2/\lambda}F\,,
\end{equation}
where $ F $ is
\begin{itemize}
	\item Non degenerate:
	\begin{equation}
	 F=\frac{1}{4}\left( 1+\frac{\lambda}{2}\right) \,;
	\end{equation}
	\item Mildly degenerate:
\begin{equation}
F=0.19 \left( 1+0.53\lambda\right) \,;
\end{equation}
	\item Degenerate:
\begin{equation}
F=\frac{1}{\pi}f(\eta)\,.
\end{equation}
\end{itemize}

\bibliographystyle{aasjournal}
\bibliography{corecollapse}

\end{document}